\documentclass[preprint]{aastex63}
\usepackage{graphicx}
\usepackage{amsmath}
\usepackage{float}
\usepackage{hyperref}

\def\iso#1#2{\mbox{${}^{#2}{\rm #1}$}}
\def\fe6#1{\iso{Fe}{6#1}}
\def\al2#1{\iso{Al}{2#1}}
\def\pu24#1{\iso{Pu}{24#1}}
\def\be1#1{\iso{Be}{1#1}}

\def\dtpar{\sigma_t}
\def\twidth{\Delta t}

\def\beq{\begin{equation}}
\def\eeq{\end{equation}}
\def\beqar{\begin{eqnarray}}
\def\eeqar{\end{eqnarray}}
\def\pfrac#1#2{\left( \frac{#1}{#2} \right)}
\def\bfrac#1#2{\left[ \frac{#1}{#2} \right]}

\def\cmin{$C_{\rm{min}}$ }

\begin{document}

\title{Supernova Dust Evolution Probed by Deep-sea \fe60 Time History}

\author{Adrienne F. Ertel}
\affil{Department of Astronomy, University of Illinois, Urbana, IL 61801, USA}
\affil{Illinois Center for the Advanced Study of the Universe, University of Illinois, Urbana, IL 61801}
\author{Brian J. Fry}
\affil{Department of Physics and Meteorology, United States Air Force Academy, Colorado Springs, CO, 80840, USA}
\author{Brian D. Fields}
\affil{Department of Astronomy, University of Illinois, Urbana, IL 61801, USA}
\affil{Department of Physics, University of Illinois, Urbana, IL 61801}
\affil{Illinois Center for the Advanced Study of the Universe, University of Illinois, Urbana IL 61801}
\author{John Ellis}
\affil{Theoretical Physics and Cosmology Group, Department of Physics, King's College London, London WC2R 2LS, UK}
\affil{NICPB, R\"avala pst.~10, 10143 Tallinn, Estonia}
\affil{Theoretical Physics Department, CERN, CH-1211 Geneva 23, Switzerland}

\begin{abstract}
  There is a wealth of data on live, undecayed \fe60 ($t_{1/2} = 2.6 \ \rm Myr$) in deep-sea deposits, the lunar regolith, cosmic rays, and Antarctic snow, which is interpreted as originating from the recent explosions of at least two near-Earth supernovae.  We use the \fe60 profiles in deep-sea sediments to estimate the timescale of supernova debris deposition beginning $\sim 3$~Myr ago.  The available data admits a variety of different profile functions, but in all cases the best-fit \fe60 pulse durations are $>1.6$~Myr when all the data is combined.  This timescale far exceeds the $\lesssim 0.1$~Myr pulse that would be expected if \fe60 was entrained in the supernova blast wave plasma.  We interpret the long signal duration as evidence that \fe60 arrives in the form of supernova dust, whose dynamics are separate from but coupled to the evolution of the blast plasma.  In this framework, the $>1.6$~Myr is that for dust stopping due to drag forces.  This scenario is consistent with the simulations in \citet{fry2020}, where the dust is magnetically trapped in supernova remnants and thereby confined around regions of the remnant dominated by supernova ejects, where magnetic fields are low.  This picture fits naturally with models of cosmic-ray injection of refractory elements as sputtered supernova dust grains and implies that the recent \fe60 detections in cosmic rays complement the fragments of grains that survived to arrive on the Earth and Moon.  Finally, we present possible tests for this scenario.  \\
  ~~\\
{KCL-PH-TH/2022-27, CERN-TH-2022-084}
\end{abstract}


\section{Introduction} \label{sec:intro}

\citet{ellis1996} suggested that deposits of live
radioactive isotopes including \fe60 could be a telltale sign for a recent near-Earth supernova explosion. 
Around the same time, \citet{korschinek1996} proposed that the high sensitivity of accelerator mass spectrometry (AMS)
could reach the levels needed to see a supernova signal.
AMS has subsequently enabled
widespread detections of live, i.e., undecayed, radioactive \fe60 in deep-sea samples from around the world ~\citep{knie1999,knie2004,fitoussi2008,ludwig2016,wallner2016,wallner2021}, which provide compelling evidence that radioisotopes from an astrophysical event reached Earth $\sim 3$~Myr ago (Mya). In addition, \fe60 has also been found in lunar samples~\citep{fimiani2016}, in cosmic rays~\citep{binns2016}, in Antarctic snow \citep{koll2019}, and in a deep-sea ferromanganese (FeMn) crust from 6 to 7~Mya~\citep{wallner2021}. These signals far exceed known terrestrial and meteoritic backgrounds. The half-life of \fe60 
$t_{1/2} = 2.60 \pm 0.05 \ \rm Myr$ \citep{rugel2009,wallner2015a,ostdiek2017} is much less than the age of the Earth, which implies that the astrophysical sources of these radioisotope deposits were relatively recent.

The explosion of at least one near-Earth supernova has been the
general interpretation of the \fe60 data ever
since the pioneering detections of \citet{knie1999},
with a distance estimated to be in the range of tens of parsecs
\citep{fields1999,fields2008}.
\citet{fry2015} expanded this analysis to consider all known
or proposed astrophysical \fe60 sources,
concluding that the \fe60 abundance and its implied distance
rule out all but core-collapse supernovae and
asymptotic giant branch (AGB) stars, as further discussed below.\footnote{It has recently
been suggested that the deposition of \fe60 $\sim 3$~Mya might have occurred as the solar system
passed through the heart of a large, dense, cold gas cloud \citep{opher2022}.
The Local Leo Cold Cloud \citep{peek2011,gry2017} was suggested as a possible target,
but the uncertainties in its kinematics, distance, and physical size make it hard to assess
the chances of collision.}

The blast from a supernova at such a distance does not itself
penetrate the heliosphere as far as the Earth's orbit
at 1 au \citep{fields2008,miller2022},
but supernova ejecta in the form of dust grains
\citep{benitez2002}
can reach the Earth and Moon \citep{athanassiadou2011,fry2016}.
Iron is one of the most refractory elements, i.e., it has a high
condensation temperature and readily forms dust, and \fe60 would be delivered
in whatever iron-bearing dust particles survive the
journey to the solar system.
\citet{fry2016} used the \fe60 flux to infer the distance
to the supernova, finding $D_{\rm SN} \sim 30-150 \ \rm pc$.
The uncertainty is large, but encouragingly, this is
precisely the plausible astrophysical range, neither so close as
to cause a mass extinction, nor so far that the supernova
material cannot reach us.
\citet{fry2016} also showed that the flux from 
the supernova blast, i.e., the gas, declines
from an initial peak, corresponding to the passage
of the dense supernova shell.
At the distances implied by the strength of the \fe60 signal, 
the duration of the blast flux peak was found to be 
at most $\sim 0.1 \ \rm Myr$.

Deep-sea sediments offer unprecedented time resolution of the \fe60 signal
at the level of a few kiloyears (kyr), opening a new window on the possible nearby supernova(e).
\citet{fitoussi2008} pioneered this approach, searching for \fe60 in a sediment using 
AMS with lower sensitivity than
is now available.  They found no evidence for the short $\sim $ few kyr signal they expected, but
showed that a potential signal emerged with time bins stretching to $\sim 1 \ \rm Myr$.
As we show below, subsequent high-sensitivity data from multiple sites and groups
 confirm that the width of the \fe60 deposition pulse arriving $\sim 3$~Myr ago exceeds
$1 \ \rm Myr$.
This timescale is much longer than
that of a blast from a single explosion \citep{fry2015},
and understanding this long timescale for \fe60 deposition
is the goal of this paper.

In this paper, we present an analysis of the \fe60 flux history for the four
well-measured deep-sea sediment cores, two
from \citet{ludwig2016} and two from \citet{wallner2016}. We develop a statistical methodology appropriate for the \fe60 data, which are dominated by counting statistics, and use this to fit the \fe60 flux for the different cores individually and in a global fit with 
a focus on the signal timescale.
We compare a variety of simple fitting functions,
and all show that the timescale must exceed 1 Myr.  We also test the ability of the data to discriminate among different time histories, finding that this is not possible with current data.

We interpret the long \fe60 deposition timescale
based on the assumption that the supernova dust
is decoupled from the gas, with different dynamics, 
so that the dust particle density profile is different
from the blast profile. Such decoupling was found by
\citet{fry2020} in a study of dust propagation in
a supernova remnant.
Here we extract the key physics of this process and
present a model for the dust flux versus time,
which we compare with the available data.
We then discuss a number of consequences and tests of our model.

Our work builds on the insights and analysis of \citet{ellison1997},
who noted that supernova grains are charged,
and that they decouple from the gas.
These authors further proposed that grains
are {\em accelerated} by the same diffusive shock acceleration processes 
that lead to cosmic-ray acceleration. Indeed, they proposed
that the sputtered atoms of the accelerated dust
are injected as cosmic rays, and that this population is responsible 
for the observed enhancement of refractory elements
in cosmic rays \citep{meyer1997}.\footnote{The idea that supernovae might accelerate dust grains goes back to Spitzer's proposal that light pressure from the explosion could accelerate surrounding {\em preexisting} interstellar dust, and possibly even be the source of heavy elements in the cosmic rays \citep{spitzer1949,wolfe1950}.
Subsequently several authors have studied grain acceleration by supernovae or other processes
\citep{hayakawa1972,wickramasinghe1974,hoang2015}, and even considered the possibility that the ultra-high-energy cosmic rays could be relativistic grains.}
\citet{giacalone2009} have performed simulations
of dust in the presence of supernova shocks, and
found that grains initially at rest were accelerated to more than 10 times the shock speed.
Our model elaborates this picture: as described below in \S\ref{sec:pinball},
we propose that the
\fe60 deposits on the Earth and Moon arise from the portions of
iron-bearing supernova dust that have survived propagation
to the Earth, while the \fe60 detected in cosmic rays \citep{binns2016} represents
the portion that was sputtered along the way.

This paper is structured as follows.  We discuss the time-resolved \fe60 sediment data in \S\ref{sec:data_intro}.  We perform fits to the data in \S\ref{sec:fits}, deriving constraints in the \fe60 deposition timescale and finding it to be $> 1 \ \rm Myr$.  We show in \S\ref{sec:pinball} that this long timescale is consistent with a picture of charged supernova dust propagation in a magnetized interstellar medium (ISM).  
We propose tests of this model in \S\ref{sec:tests}. We summarize our conclusions in \S\ref{sec:conc}.

\section{Time-resolved Measurements of \texorpdfstring{\fe60}{60Fe} Deposition on Earth} \label{sec:data_intro}

The evidence for \fe60 deposition on Earth comes mainly from deep-sea deposits,
namely ferromanganese (FeMn) crusts and nodules, as well as sediment cores. 
The FeMn crusts and nodules exhibit relatively slow growth, $\sim$~few~mm Myr$^{-1}$, which implies less dilution of the small extraterrestrial signal and facilitated the first detections \citep{knie1999}.  However, the slow growth rate makes it more difficult to obtain good time resolution.  On the other hand, the growth (i.e., sedimentation) rates of deep-sea sediments are typically about a factor of 1000 faster, namely  $\sim$~few~mm kyr$^{-1}$. This means that a larger sample and more processing are needed to find the signal, but it is also easier to obtain good time resolution. 

The importance of the \fe60 signal width as an observable goes back at least to \citet{feige2014}.  She illustrated possible pulse shapes assuming a Gaussian form, emphasizing the trade-off between ability to resolve the width and dilution of the signal at its peak.  When \citet{fitoussi2008} performed the first sediment measurements with relatively low \fe60 sensitivity, their results were hampered by this trade-off, which led to small \fe60 counts spread over $\sim 1  \ \rm Myr$.  In addition, they adopted time bins of about 10 kyr, anticipating a short signal consistent with a Sedov-Taylor blast; this further diluted the signal.  By attaining improved sensitivity and adopting larger time bins, \citet{ludwig2016} and \citet{wallner2016} later unambiguously resolved the signal; \citet{feige2018} performed an initial Gaussian fit to the binned \citet{wallner2016} sediment data.  The time is ripe for a detailed joint analysis of these results, as presented in this paper.

\begin{figure}[!hb] 
	\centering
  \includegraphics[width=15cm]{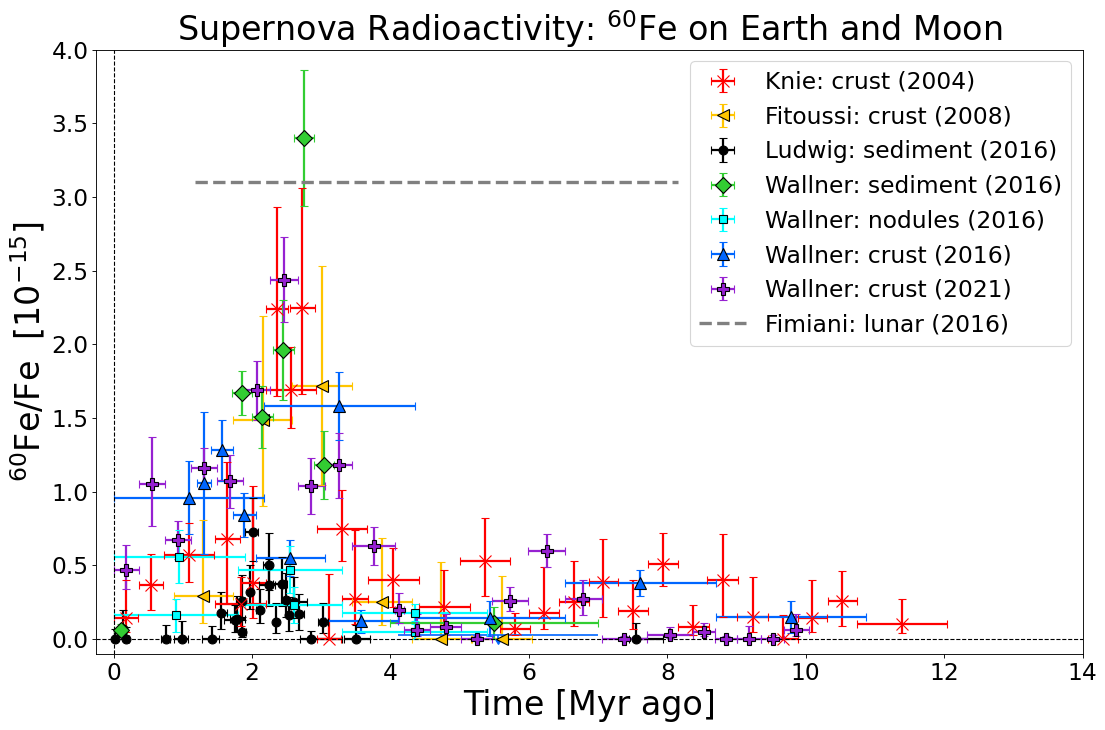}
  \caption{\textit{Terrestrial and lunar detections of $^{60}$Fe.} The $^{60}$Fe/Fe fractions found in deep-sea sediments, FeMn crusts, and Apollo lunar samples; the data are not decay corrected for \fe60; however the \citet{knie2004} and \citet{fitoussi2008} times have been updated to take account of the latest \be10 half-life ($t_{1/2} = 1.387 \pm 0.012$ Myr; \citep{korschinek2010}).  None of the data are background corrected.  All of the data with time resolution show a signal around $\sim  2-3$~Mya. The amplitude differences may reflect variations in iron uptake, latitude variations in iron deposition, and/or differences in sampling technique.  Note the appearance of a second distinct peak around $\sim  6-7$~Mya in the \citet{wallner2021} data.}
  \label{fig:data}
\end{figure}

\newpage 

\subsection{Measurements of Pulses around 3 and 7 Myr Ago}

The presence of deep-sea pulses of live \fe60 is now compelling, providing strong evidence for recent nearby supernovae.
Figure \ref{fig:data} compiles the published \fe60 data. The figure shows the detected \fe60/Fe isotope fraction versus time, categorized by both sample type and research group; none of the data have been background corrected. The data show two distinct peaks, with all groups agreeing on a \fe60 peak around $2-3$ Mya and the \citet{wallner2016,wallner2021} data indicating another peak around $6-7$ Mya.  Note that this peak at 7 Mya only appears clearly in the \citet{wallner2021} data, whose \fe60 machine background has finally been lowered enough to show a distinct peak; the background in earlier efforts obscured the signal.  Not included in this figure are the recent \fe60 flux measurements by \citet{koll2019} and \citet{wallner2020}, as they cover collectively only the last 30 kyr and would not be visibly distinguishable from the origin (see \S \ref{subsec:infall}). The \citet{fimiani2016} lunar data are included for completeness only: due to micrometeorite gardening effects on the lunar regolith, the data cannot be time resolved to better than $\sim 8 \ \rm  Myr$.\footnote{\fe60/Fe ratios were not quoted explicitly in the \citet{fimiani2016} paper: we calculated an average value using the \fe60 concentration values from their Figure 3 and the Fe concentration from their Table 2 for the relevant samples (1,2,4,5,6,7,9,10).  The average is $\fe60$/Fe $\sim 3.1 \times 10^{-15}$.}  Because of this time range, the lunar data include the signals due to all supernovae in the last 10 Myr; however, due to the half-life of \fe60, the contribution of \fe60 from the pulse at 7 Mya is only around 10\%. 

When comparing the \fe60/Fe measurements, one should bear in mind that the geographical distribution of \fe60 may not be uniform, and that the uptake, $U$, of \fe60 varies in different materials.  Specifically, it was shown in \citet{fry2016} that the transport of \fe60 through the atmosphere should not be isotropic, but rather would favor middle latitudes $\sim 60^\circ$, with minima at the equator and poles \citep[see also][]{dhomse2013}, yielding a factor of $\sim 3$  in the global difference. This may explain some of the large difference between the flux values reported by \citet{wallner2016} taken at $\sim 38^\circ \rm S$ versus the those in the data of \citet{ludwig2016} taken at $3^\circ \rm S$. In addition, ocean currents may cause variations with longitude in the deposition of \fe60 in FeMn crusts and deep-sea sediments at similar latitudes. Differences in analytical technique and sample processing can also affect the final result.
We note finally that the uptake is expected to be 100\% for sediments and snow, whereas the uptake in FeMn crusts is subject to considerable uncertainty and debate, and may vary depending on location and local conditions \citep{bishop2011}.

In this paper, we focus on the sediment data when estimating the timescale of astrophysical \fe60 deposition, in view of the availability of multiple samples with time resolution from deep-sea sediments \citep{ludwig2016, wallner2016}. We note that the \citet{wallner2021} \fe60 data from the FeMn crust also has excellent time resolution compared to earlier studies. However, analysis of the deposition timescale requires strict accounting for geophysical processes that might disturb the signal, and this is more straightforward when the data are from the same sample type.  By focusing on the unbinned sediment data, we can control most of the variables between the two data sets and therefore make fair comparisons.  It should be noted that this approach by necessity deals with very small-count statistics; we address the issue in \S\ref{subsec:flux}. We look forward to additional measurements of the 7 Myr peak, ideally in sediments, to allow for a similar analysis of that event. 

\subsection{Recent \texorpdfstring{\fe60}{60Fe} infall} \label{subsec:infall}

\citet{koll2019} and \citet{wallner2020} have shown independently that there is recent and ongoing infall of extrasolar \fe60 onto Earth.
\citet{koll2019} detected an \fe60 signal in Antarctic snow deposited over the last 20 yr and use isotopic ratios to show it is not from meteoritic material and must therefore be extrasolar.  They suggest that the signal is due to the solar system passing through the Local Interstellar Cloud (a nearby higher-density region of the Local Bubble).  \citet{wallner2020} also detected a fairly steady \fe60 signal in deep-sea sediments over the last 33 kyr, in line with the \citet{koll2019} detection. However, they do not see the sharp increase in the \fe60 signal that would be expected from the solar system entering the Local Interstellar Cloud. There are other possibilities for the persistent \fe60 flux, 
including continued  delivery of dust following the most recent pulse $\sim 3$~Mya, or flux from the Earth's motion through the local ISM.  
Further analysis of \fe60 in FeMn crusts and deep-sea sediments focused on the age range from 40 Kya to 1 Mya could add significant insight into
the origin of the observed recent infall. 

That said, the main purpose of this paper is to characterize the peaks that are clearly evident 
in the data, and we do not attempt to fit the low-level \fe60 flux outside the peaks.

\section{Supernova Dust Deposition Timescale from \texorpdfstring{\fe60}{60Fe} in Deep-sea Sediments } \label{sec:fits}

Ocean sediments are the most suitable tracers for timescale analysis due to their rapid growth rate $\sim$~few~mm kyr$^{-1}$, 
which is a factor $\sim 10^3$ faster than that of the ferromanganese crusts.  This rapid growth allows the sediment column to be sampled more finely, leading to much better time resolution, but at the cost of a lower \fe60 concentration.  We study the sediment data of \citet{wallner2016} and \citet{ludwig2016}, which were taken from different ocean drilling program
cores and analyzed independently. \citet{wallner2016} made measurements in four cores, of which two probe a very limited time range, leaving two cores (4521 and 4953) with sufficient data for our analysis. The two cores (848 and 851) studied by \citet{ludwig2016} are both sufficiently well sampled for our purposes. Unfortunately, the pioneering \citet{fitoussi2008} data did not have sufficient sensitivity for a well-resolved time series; they still provide useful consistency checks, but we do not use them for our full timescale study.

Both \citet{wallner2016} and \citet{ludwig2016} used AMS measurements of \fe60 atoms in the samples, from which \fe60/Fe isotope fractions are derived, as well as the \fe60 flux and its time-integrated fluence.  Additionally, both groups studied blank samples to infer that their background is negligible, and therefore all of the \fe60 counts are significant.

\begin{table}[htb]
    \centering
    \caption{Summary of Timescales of Nonzero Counts in Deep-sea Drill Cores
    \label{tab:data-timespan}
    }
    \begin{tabular}{c|cc|c}
    \hline\hline
    Core & $t_{\rm last}$ & $t_{\rm first}$ & $\delta t = t_{\rm first}-t_{\rm last}$   \\
    Name & (Mya) & (Mya) & (Myr) \\
    \hline
     Ludwig 848 & 1.528 & 2.604 & 1.076 \\
     Ludwig 851 & 1.735 & 3.045 & 1.310 \\
     Wallner 4521 & 1.78 & 2.57 & 0.79 \\
     Wallner 4953 & 1.71 & 3.18 & 1.47 \\
     \hline
    All Cores & 1.528 & 3.18 & 1.65  \\
    \hline \hline
    \end{tabular} 
\end{table}

\noindent
Before performing our fits, it is useful to note that the raw data
give useful information about the signal width.
For each core, there is a distribution of nonzero \fe60 counts. 
The interval $\delta t = t_{\rm first}-t_{\rm last}$ between the time $t_{\rm first}$ of the first nonzero count 
and the time $t_{\rm last}$ of the last nonzero count thus gives a minimum
time span for each core.  This assumes that background effects are negligible.
These results are summarized in Table \ref{tab:data-timespan},
where we see that the signal durations in the individual cores span $\delta t=0.79$ to 1.47 Myr.
It is important to note that the Ludwig cores have multiple measurements with zero counts before and after the ranges of their nonzero measured counts; this is not the case
for the Wallner data.  Thus the Ludwig results in Table \ref{tab:data-timespan} represent an estimate of their signal duration; though Poisson fluctuations or flux beneath their sensitivity could lead to a longer signal.  For the Wallner data, there are no leading and trailing zero counts, and so the results in Table \ref{tab:data-timespan} are certainly a lower limit to the duration in the cores they measured.

If we further assume there are no systematic differences in absolute timing,
then all of the cores together probe the range of the signal.
Then the global minimum time span is the interval between overall first and last nonzero counts
among all the cores. 
Globally, the \fe60 detections in sediments span 1.65 Myr.  
We already see that the signals are quite
long compared to the Sedov timescale $\lesssim 0.1$ Myr.  As we now turn to fits,
 we will find that these characteristic timescales set lower limits to our results.

\subsection{Analysis of the \texorpdfstring{\fe60}{60Fe} timescale} \label{subsec:flux}

The purpose of this work is to determine the deposition or ``raindown'' time onto Earth of \fe60 during the recent pulse
$\sim 3$~Mya, and to interpret what this timescale implies about the propagation of supernova-produced material inside the remnant.  In order to perform this analysis, we fit the observed \fe60 signal with a number of 3-parameter pulse shapes to see which one best described the data, while assuming the errors are dominated by Poisson counting statistics.  
These shapes included a Gaussian, sawtooth, reverse sawtooth (for comparison, despite our doubt that this is a physically plausible profile), 
and a symmetric triangle;
expressions appear in Appendix \ref{app:fitfuncs}.  We also performed one 4-parameter fit with an asymmetric triangle to see if there is a 
preferred slant to the data.  

The general shape of the \fe60 data is of particular interest, because the sharpness and 
slant of the shape can provide insight into the astrophysics of the deposition of the \fe60 and its path
within the supernova remnant.  Predictions in the literature to date have assumed the \fe60 traces the gas phase of the blast.  Under this assumption, \citet{fry2015} showed that if the \fe60 is well mixed in a Sedov blast, the signal appears discontinuously with the arrival of the forward shock, and decreases thereafter from this maximum.
\citet{chaikin2022} included the effects of incomplete mixing of \fe60 in the remnant gas, and also allowed for effects of the Earth's motion.  They too found that the \fe60 pulse begins abruptly and is concentrated in a pulse, but found that the signal can linger thereafter.  Thus, these models would favor the discontinuous profiles we have considered --- the sawtooth form, or a cut exponential. As we will see, there are other possibilities if the \fe60 is in dust that is decoupled from the gas, so we have chosen a suite of different fitting functions to allow for a range of possible \fe60 flux histories.

In the published versions of their work, both \citet{wallner2016} and \citet{ludwig2016} bin their data, 
which serves to demonstrate the strength and overall peak of the signal.  For the purposes of this analysis, we use the original, unbinned data, as it removes extraneous smoothing, and we are specifically interested in fitting the unbinned shape.  The unbinned \fe60 data can be found in Table S.4 of \citet{wallner2016} and in Tables A.1 and A.2 of \citet{ludwig2015}. Both \citet{wallner2016} and \citet{ludwig2016} assume zero background for their analyses. As mentioned above, recent works by \citet{wallner2020} and \citet{koll2019} have found a nonzero \fe60 background today. This minor discrepancy is discussed in more detail in \S\ref{subsec:infall}.  For this work, we use the zero background estimate assumed by the original analyses. 

In each sediment, AMS measures individual \fe60 atoms in each sediment segment corresponding to a time bin.
The numbers of counts measured is small: in all of their sediments, \citet{ludwig2016} detected a total of 89 atoms of \fe60, while \citet{wallner2016} measured a total of 288 atoms.
The $n_i$ number of \fe60 counts in each time bin $i$ is therefore also small, with $n_i \le 23$
and often $n_i < 10$.
As a result, the Poisson errors in the counts dominate the uncertainties in the resulting \fe60 flux.
We therefore tailor our analysis to identify the dependence on count numbers $n_i$ and to treat these Poisson errors appropriately. 

The observed \fe60 flux values $\Phi_{60}$ (the deposition into the material) are computed by combining 
the measured \fe60 counts with properties of the sediment, as follows.  
For each time $t_i$, 
a number $n_i$ of \fe60 atoms are measured.
From this, the
observed \fe60/Fe ratio is
determined:
\begin{equation}
\label{eq:r60}
\pfrac{\fe60}{{\rm Fe}}_i = s_i \ n_i \, ,   \end{equation}
where
the scaling $s_i$ is due to variations in efficiency  and
other factors, and unique
for each measurement. 
We infer the scalings 
from the reported $n_i$ and $(\fe60/\mathrm{Fe})_i$.
In the cases of null measurements,
we follow the same procedure as both experimental groups, using the
\citet{feldman1998} 
prescription for calculating 69.29\% CL limits
based on
$n_i = 0$ counts, and a background $b=0$.
This gives an effective limit
$n_i^{\rm eff} < 1.29$, which
we use to infer the scaling $s_i$
for these measurements.
 
We use the following equations to correct the data for decays:
\begin{align} \label{eq:decay_cor}
    \pfrac{\fe60}{\rm Fe}_{\rm c} &= \pfrac{\fe60}{\rm Fe}\, \exp{\bfrac{t_{i}}{\tau}} \, , \\
    \sigma_{60_{\rm c}} &= \sqrt{ \exp{\bfrac{2\,t_{i}}{\tau}} \sigma_{60}^2 + \left(\dfrac{\fe60}{\mathrm{Fe}} \ \dfrac{1}{\tau}\right)^{2} \exp{\bfrac{2\,t_{i}}{\tau}} \sigma_i^2} \, .
\end{align}
Here and throughout, the subscript ``${\rm c}$'' indicates that the quantity is decay corrected,
$t_{i}$ is the observed time with uncertainty $\sigma_i$, $\tau$ is the mean lifetime of \fe60, and $\sigma_{60_{\rm c}}$ is the uncertainty in the decay-corrected  \fe60/Fe ratio.
To calculate the \fe60 flux, we use
\begin{align} 
\label{eq:flux}
    \Phi_{\rm 60c} &= n_{\rm 60c} \dot{h} =  \pfrac{\fe60}{\rm Fe}_{\rm c} n_{\rm Fe} \, \dot{h} = \frac{X_{\rm Fe}}{A_{\rm Fe} m_{\rm u}} \pfrac{\fe60}{\rm Fe}_{\rm c} \, \rho \, \dot{h} \\
        &= c_{\mathrm{Fe}} \, \pfrac{\fe60}{\rm Fe}_{\rm c} \, \rho \, \dot{h} \, , \\
        \label{eq:fluxsig}
    \sigma_{\Phi} &= c_{\mathrm{Fe}} \, \sigma_{60c}  \, \rho \, \dot{h} \, . 
\end{align}
 In eq.~(\ref{eq:flux}),  $\Phi$ is the flux of \fe60, and
 $n_{60}$ is the \fe60 number density.
 The sediment mass density is $\rho$, $\dot{h}$ is the sedimentation rate,
 and $X_{\rm Fe} = \rho_{\rm Fe}/\rho$
 is the mass fraction of iron in the sediment.  
The factor $c_{\mathrm{Fe}} = X_{\rm Fe}/A_{\rm Fe} m_{\rm u}$ measures 
the concentration of iron in the sediment in atoms per unit mass, 
with $A_{\rm Fe}$ as the mean molecular weight of iron and $m_{\rm u}$ as the atomic mass unit.

Due to the format of the data provided in the relevant papers, the conversion of the \fe60/Fe ratio into a flux was 
by necessity different for the two groups.  \citet{wallner2016} calculated the flux in their Table S.4, 
following the formula in Eq.~(\ref{eq:flux}), and we use those numbers directly. However,
\citet{ludwig2016} used a protocol in their chemical sample treatment
(citrate-bicarbonate-dithionite, hereafter CBD) that effectively separates larger iron-bearing grains from
smaller grains, which they argue should arise from fossilized magnetotatic bacteria.
Their AMS measurements thus give a $(\fe60/{\rm Fe})_{\rm CBD}$ ratio for this material, and thus their scaling $c_{\rm Fe}$ of iron per mass in the bulk unprocessed sediment includes a factor $Y_{\rm CBD}$ for the fraction of iron selected by the CBD process.
This implicitly assumes that the Fe-bearing material excluded from the CBD protocol does not contain \fe60.
For the flux calculation in Eq.~(\ref{eq:flux}), the unbinned \fe60/Fe ratio and extracted iron are from Tables A.1 and A.2 in \citet{ludwig2015} (the binned versions of which appear in~\citet{ludwig2016}), 
the sediment density is taken from Table 6.3 in \citet{ludwig2015}, 
and the sedimentation rate is from Figure 1 in~\citet{ludwig2016}.\footnote{We are indebted to P. Ludwig for helping us understand and combine the different data sets.} 

Combining eqs.~(\ref{eq:r60})
and (\ref{eq:flux}),
we arrive at the relationship between the flux and \fe60 counts.  For time $t_i$,
we have
\begin{equation}
\label{eq:fluxscaling}
    \Phi_{60,c}(t_i)
    = \frac{X_{\rm Fe}}{A_i m_{\rm u}} 
    \rho \dot{h} \ e^{t_i/\tau} \ s_i \ n_i \ \equiv \ \varphi_i \ n_i \, .
\end{equation}
Given the counts $n_i$ and the other observables,
eq.~(\ref{eq:fluxscaling}) allows us to determine the flux scalings $\varphi_i$.
As we now see, these allow us to compare the observed
fluxes to model predictions.

Our goal is to fit the observed flux data with several simple models
that capture different qualitative trends one might expect in the \fe60
flux versus time, $\Phi_{\rm model}$.
The flux profile versus time is described by a set of parameters
$\vec{\theta}$, so that we have $\Phi(t;\vec{\theta})$.
In designing our fitting procedure, we are guided by the fact that the 
uncertainties in the \fe60 flux are dominated by the Poisson errors in the \fe60 counts,
which is the case for both the \cite{wallner2016} and \citet{ludwig2016} data sets.
We have designed our fitting procedure to accommodate this situation,
closely following the approach laid out by \citet{cash1979},
originally for determining X-ray fluxes from photon count measurements
dominated by Poisson uncertainties.

Our analysis requires that at each time $t_i$
we specify the expected number $\mu_i$
of events, based on the model flux.  To do this,  
we evaluate $\Phi_{\rm model}(t_i;\vec{\theta})$
and then infer $\mu_i(\vec{\theta}) = \Phi_{\rm model}(t_i;\vec{\theta})/\varphi_i$, 
using the the scaling $\varphi_i$ found in eq.~(\ref{eq:fluxscaling}).
For each measured number of counts
$n_i$ the fit function with parameters $\vec{\theta}$ gives an expected value $\mu_i(\vec{\theta})$.

We now construct a Poisson-based likelihood for the fit given
the data.
For time $t_i$,
the likelihood for the fit given the data
is just the Poisson probability
${\cal L}_i(n_i|\vec{\theta}) = P(n_i|\mu_i)$,
where $P(n|\mu) = \mu^n e^{\mu}/n!$ is the Poisson probability of $n$ counts given a mean $\mu$.
The total probability for the fit given all of the data
is just the product of the likelihoods at each time:
\begin{equation}
\label{eq:likelihood}
\mathcal{L}(\mbox{data}|\vec{\theta}) = \prod_i {\cal L}_i(n_i|\mu_i) 
= \prod_i \frac{\mu_i^{n_i}}{n_i!} e^{-\mu_i}
\end{equation}
where the fit parameter dependence is through $\mu_i(\vec{\theta})$.
The negative of the logarithm of the total
likelihood in eq.~(\ref{eq:likelihood}) is 
\begin{equation}
\label{eq:poisson}
C(\vec{\theta}) = - \ln \mathcal{L}(\mbox{data}|\vec{\theta})
= - \sum_i \left( \mu_i - n_i \ln \mu_i \right)
+ \mbox{const} \, .
\end{equation}
Here the constant sums
terms with $\ln n_i$ dependencies that do not depend on the fit parameters $\vec{\theta}$, which
means that it does not affect the relative likelihoods of different parameter choices, so we follow the usual practice and neglect it.
The function $C$ thus depends on the data 
and the fit parameters, and determines the goodness of fit in a way closely analogous to the role of a $\chi^2$ for data that is continuous rather than discrete.

For a given dataset and fitting function $\Phi_{\rm model}(t;\vec{\theta})$,
eq.~(\ref{eq:poisson}) will give different values to  $C$ for different choices of the parameters $\vec{\theta}$.
The likelihood is maximized at the minimum value for $C$, which we call
$C_{\rm min}$, and which we will use to assess goodness of fit. 
The parameters $\vec{\theta}$ giving \cmin 
are the best-fit values.
We use Monte Carlo methods to explore the parameter space, find the best-fit parameters,
and characterize their uncertainties.

We note that the physical picture of supernova radioisotope deposition could 
include an abrupt onset to the flux.  This could occur
if \fe60 is entrained in a blast wave, so that  the onset of the \fe60 coincides with
the forward shock's arrival at the solar system.
To allow for this, we consider some fitting functions where the 
flux has an abrupt onset and/or halt.   In these cases, there are times before
or after the blast passage, for which there is no signal: $\mu_i = 0$, 
which means that the expected number of counts must be zero.  
If the measured number of counts is nonzero for these times, then the Poisson probability is zero
($C \rightarrow - \infty$)
and this set of fit parameters is completely ruled out.  In other words, our method
automatically rejects the models (regions of parameter space) that predict no counts where some are observed. 
On the other hand, the converse is not true:  if the fit has
$\mu_i > 0$, Poisson statistics allow for cases where $n_i=0$, albeit with a penalty.

Each of the four sediment cores was fitted separately, as each core is a separate time column, and in many cases, data points 
from different cores lie on the same time slice, which creates difficulties with Poisson statistics.  We were also interested 
in examining any differences we could find in the pulse arrival time and the peak pulse time, to see if there were differences 
in the timing calibration between the different sediments.  In order to ensure a universal resolution for all the pulse shapes, 
we enforced the same initial parameters across all of the 3-parameter fits (similar values were used for the 4-parameter fit, 
although it is not statistically comparable).  

To test our methodology, we generated 
simulated data points, drawn from a predetermined
flux history $\Phi_{\rm true}(t)$.
We randomly chose sample times $t_i$,
and drew counts $n_i$ from a Poisson distribution appropriate for our flux history.
We found that our method generally performed well:  the best-fit parameters were 
close to 
the parameters of the known input 
$\Phi_{\rm true}$.  However, we did find that the accuracy
and precision of the width
and thus timescale parameter depends on the functional form and the time distributions of measurements.
In particular, we found that a crucial factor is
the number of measured points with zero counts before and after the bulk of the signal.  
The more leading and trailing null points, the better the width was determined.  On the other hand, if there were only one or two null points on either side of the signal, the width was less well constrained, with the true width being at
the low end of the range allowed by the fit.
This is a manifestation of the physical effect that in noisy data with small mean numbers of counts, one or two bins with zero counts do
not strongly constrain the fit; rather, many are needed to exclude the models that span wide timescales.   We will see this behavior manifested in
the fits below.

\subsection{Results of Fits} \label{subsec:fit_results}

To allow for a variety of possible \fe60 time histories,
we fit the flux data with six possible 3-parameter shapes, chosen for mathematical simplicity and resemblance to
physically  motivated trends suggested in the literature. 
Their mathematical expressions are given in in Appendix \ref{app:fitfuncs}.
Three of these give a signal that has a finite
duration:  (1) a symmetric triangle with equal duration linear rise and fall around a peak,  
(2) a sawtooth that begins abruptly at a peak and falls linearly to zero,
(3) and a reverse sawtooth that rises linearly  from zero to a peak, then drops to zero.
We explored three additional profiles that allow comparison to traditional fits and explore a more gradual rise and fall that formally  never goes to zero: (1) a Gaussian, (2) a Lorentzian, and (3) a cut exponential that starts at a peak and then drops exponentially.  

\begin{figure}[htbp]  
	\centering
  \includegraphics[width=18cm]{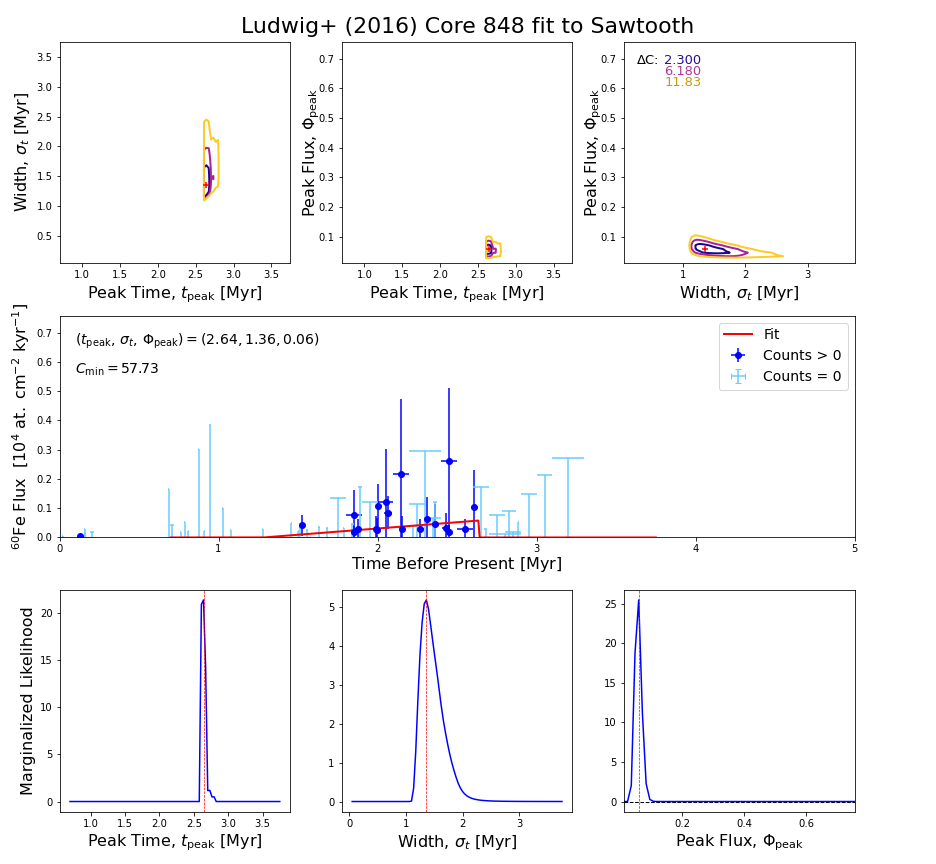} 
  \caption{\textit{Sawtooth fits to the terrestrial $^{60}$Fe flux of \cite{ludwig2016} core 848.}  The flux is in units of $[10^4 \ \rm atoms \ cm^{-2} \ kyr^{-1}]$ throughout. \textbf{Upper plots:} the confidence intervals as contours for the three fit parameters (peak time, signal width, and peak flux).  \textbf{Middle plot:} the \fe60 flux vs. time, with the sawtooth best fit overlaid in red.  The dark blue points show the calculated flux for the detected \fe60 counts, while the light blue points show the flux upper limits for the nondetections.  Error bars in both flux and time are included, although the time errors can be smaller than the data point itself.  In the upper left corner are listed the best-fit peak time, width, and peak flux values for the sawtooth fit, as well as the $C_{\mathrm{min}}$ parameter for the fit.  \textbf{Lower plots:} the marginalized likelihood for each of the three parameters, with the best-fit value indicated by a red line.}
  \label{fig:sawtooth_L848}
\end{figure}

\begin{figure}[htbp] 	
	\centering
  \includegraphics[width=18cm]{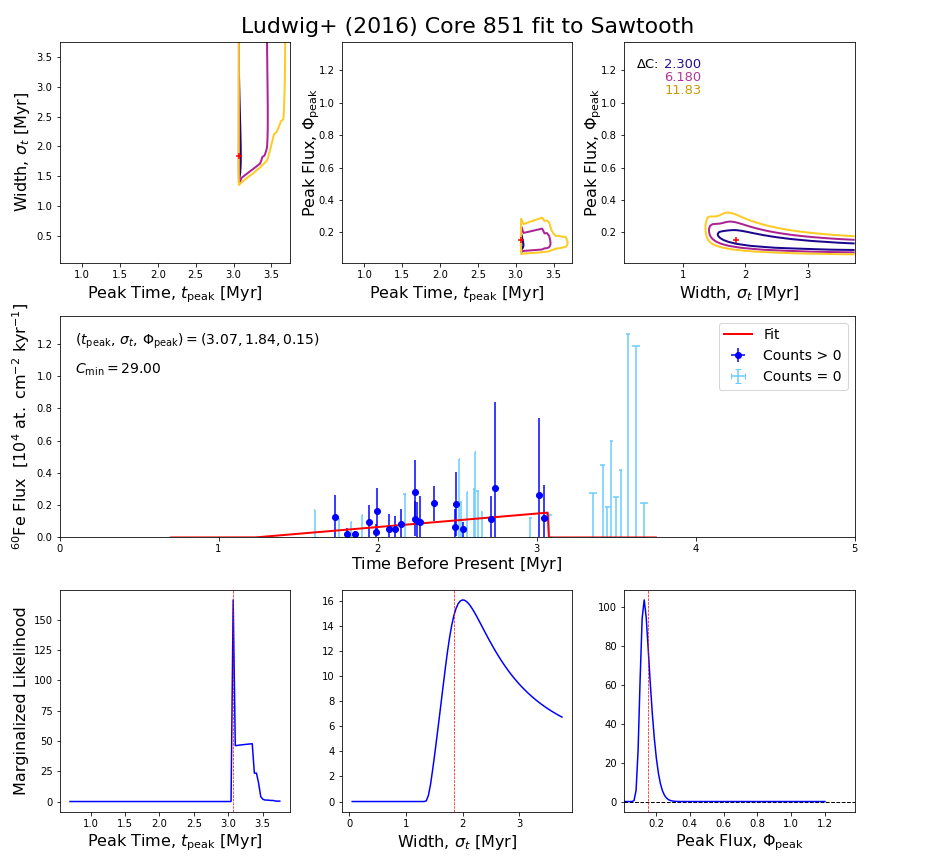} 
  \caption{\it Sawtooth fits to the terrestrial $^{60}$Fe flux of \cite{ludwig2016} core 851, as in Fig. \ref{fig:sawtooth_L848}.}
  \label{fig:sawtooth_L851}
\end{figure}

\begin{figure}[htbp]  	
	\centering
  \includegraphics[width=18cm]{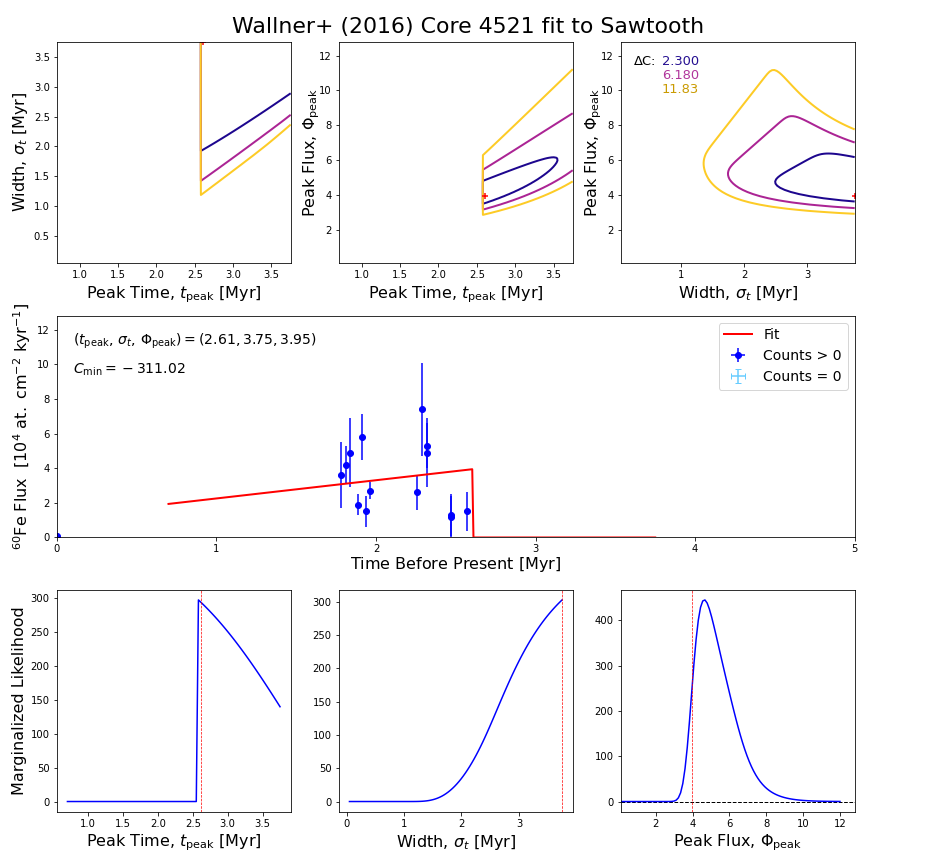} 
  \caption{\it Sawtooth fits to the terrestrial $^{60}$Fe flux of \cite{wallner2016} core 4521, as in Fig. \ref{fig:sawtooth_L848}.}
  \label{fig:sawtooth_W4521}
\end{figure}

\begin{figure}[htbp]  	
	\centering
  \includegraphics[width=18cm]{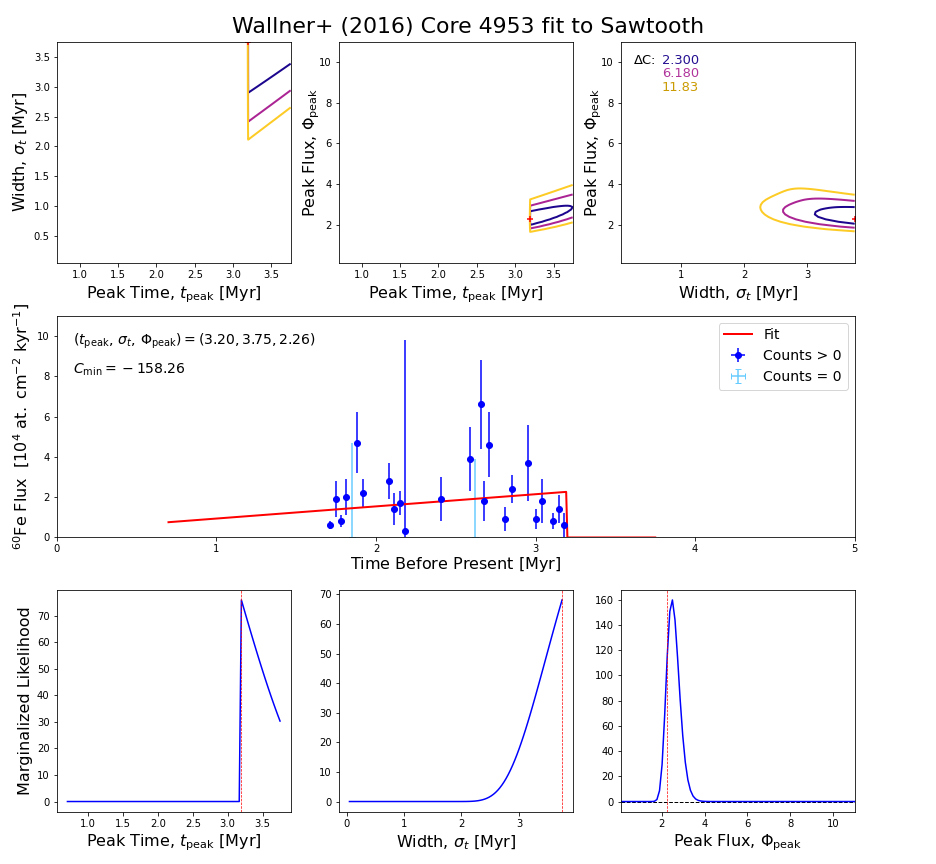} 
  \caption{\it Sawtooth fits to the terrestrial $^{60}$Fe flux of \cite{wallner2016} core 4953, as in Fig. \ref{fig:sawtooth_L848}.}
  \label{fig:sawtooth_W4953}
\end{figure}

\begin{figure}[htbp] 	
	\centering
  \includegraphics[width=18cm]{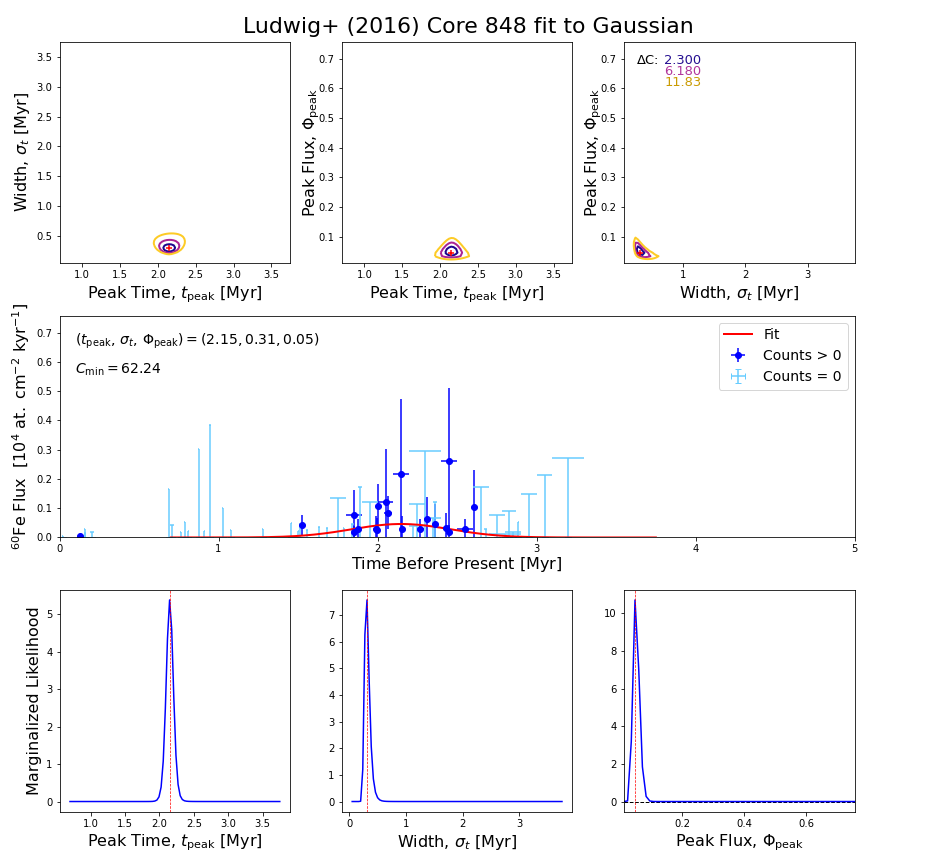} 
  \caption{\it Gaussian fits to the terrestrial $^{60}$Fe flux, \cite{ludwig2016} core 848, as in Fig. \ref{fig:sawtooth_L848}.  Shown here as an example of an alternative fit.}
  \label{fig:gaussian_L848}
\end{figure}

The data for each sediment core were fit with each specific fit model.  The results are summarized in Tables \ref{tab:cmin} and \ref{tab:fwhm}, but for brevity we plot results only for select cases.  Figures \ref{fig:sawtooth_L848}--\ref{fig:sawtooth_W4953} show results for all cores using the sawtooth fit, while Fig.~\ref{fig:gaussian_L848} shows the results for the Ludwig 848 core using a Gaussian fit.   Each fit figure shows the flux versus time data for a single core.  The middle part of the figure plots the \fe60 flux (in $10^4 \, \rm atoms \ cm^{-2} \ kyr^{-1}$) versus time (in Mya), including both detections and nondetections.  Overlaid on top of these points is the best-fit curve, with the \cmin statistic and the fit's peak time ($t_{\rm{peak}}$), peak flux ($\Phi_{\rm{peak}}$), and width ($\sigma_t$) given in the upper left corner.  The top three plots in each figure display two-dimensional projections of the three parameters of each fit model, exhibiting the contour confidence levels of $\Delta C$ = $C-C_{\rm min}$ corresponding to 1$\sigma$, 2$\sigma$, and 3$\sigma$ for a three-dimensional Gaussian;  the bottom three plots show the marginalized likelihood for each parameter normalized so that the peak is at 1.

\begin{figure}[htbp!] 	
	\centering
  \includegraphics[width=18cm]{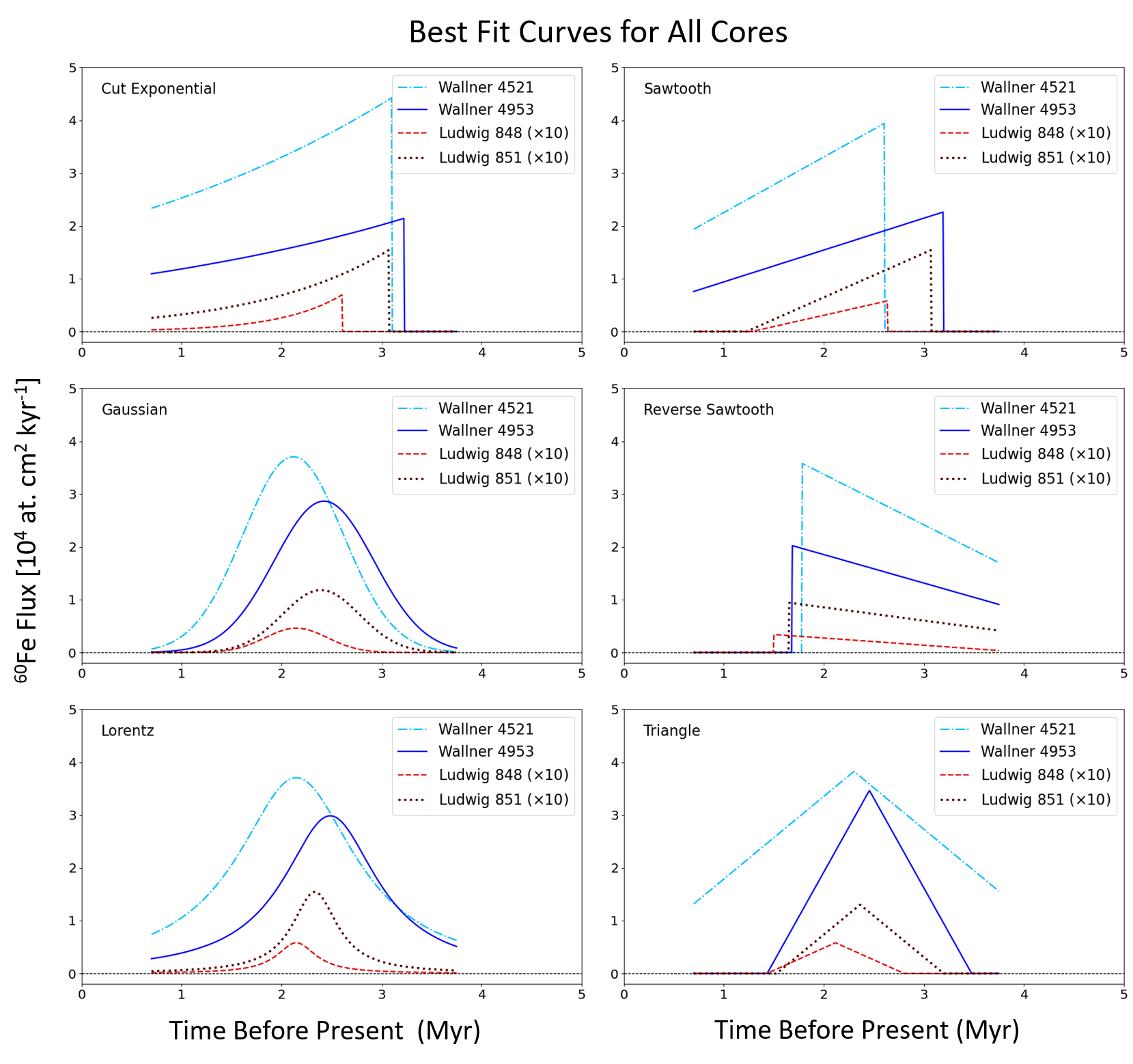}
  \caption{\it Best-fit curves for each 3-parameter fit for each core. The Ludwig fluxes have been multiplied by 10 for ease of comparison.  Curves are plotted over the time domain for which there is data from at least one measurement.
  \label{fig:all_fits}}
\end{figure}

\begin{figure}[htbp] 
	\centering
  \includegraphics[width=15cm]{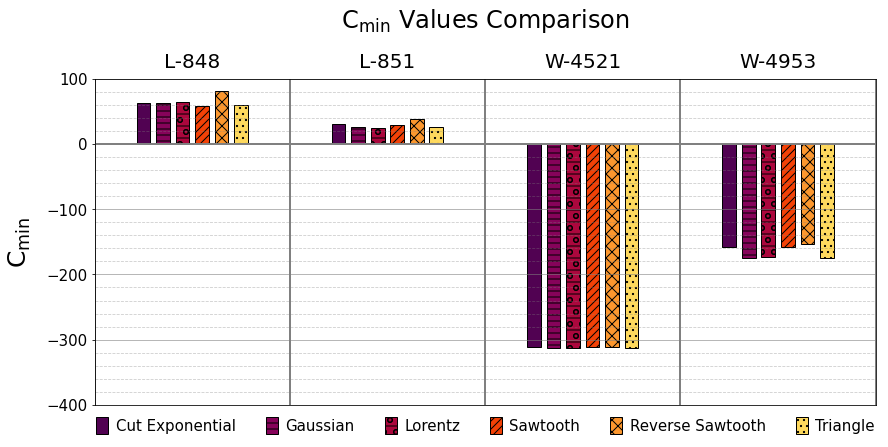}\\
  \caption{\textit{Assessing goodness of fit:  3-parameter fit \cmin comparisons for each of the 6 fits per core.}  The lower (more negative) the \cmin value, the better the fit.  Note that, while the fits can be compared with each other for each individual core, the fits should not be compared between cores.
  We see that the best-fit shape varies between cores, with no clear global preference.}
  \label{fig:cmin}
\end{figure}

\begin{figure}[htbp] 	
	\centering
  \includegraphics[width=14cm]{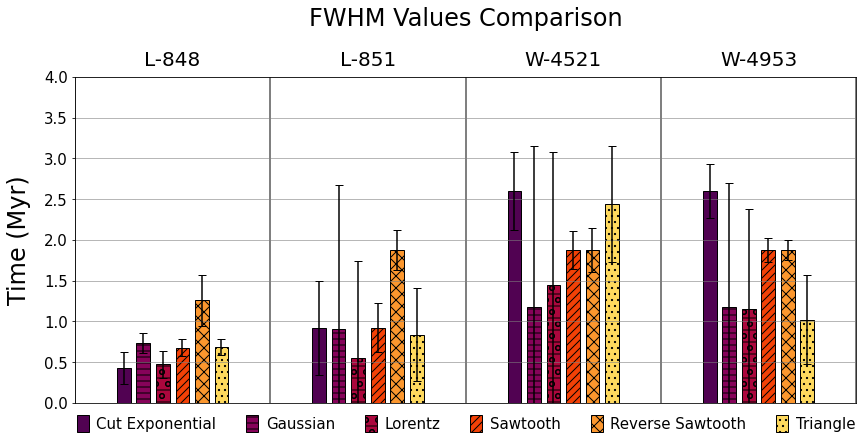}\\
  \vspace{5mm}
  \includegraphics[width=14cm]{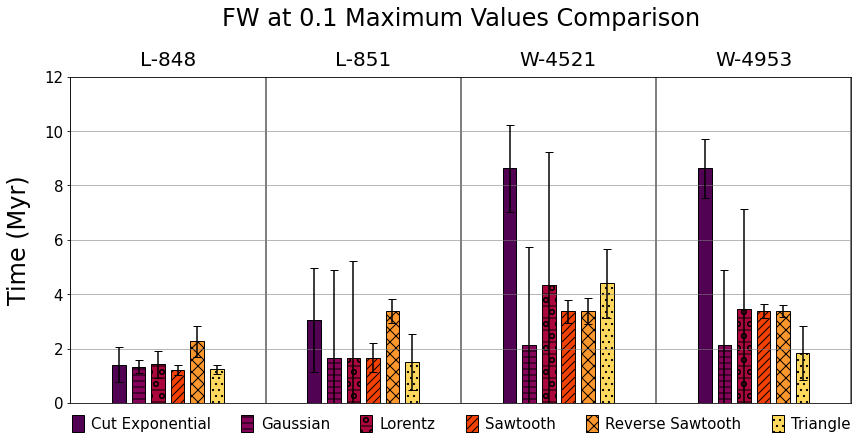}
  \caption{\textit{3-parameter fit time width comparisons.}  
  \textbf{ Top:} FWHM calculated for each of the 6 fits per core. \textbf{Bottom:} FW at 0.1 maximum height, which is closer to the true overall timescale. The errors shown are 
  the uncertainties in the parameters: they do not reflect the goodness of the fits. A small error on a best-fit value does not indicate that this particular shape is the preferred pulse.  Instead, the goodness of fit is determined by the Poisson \cmin value \citep{cash1979}.}
  \label{fig:fwhm}
\end{figure}

Figure \ref{fig:sawtooth_L848} shows the results for a sawtooth fit to the Ludwig core 848, and serves as an exemplar for similar plots of other cores and fits.  In the top panels, we see two-dimensional slices of the $\vec{\theta} = (t_{\rm peak},\dtpar,\Phi_{\rm peak})$ parameter space.  The red dot gives the location of the best fit, where $C(\vec{\theta})=C_{\rm min}$. The surrounding contours correspond
to the 68\%, 95\%, and 99\% confidence-level values.
We see that the best-fit regions are relatively compact, meaning that the best fit is well determined. The peak flux and peak time values are the best determined, with quite small uncertainties.  The width parameter shows a broader distribution and hence larger uncertainty.  
These trends are reflected in the bottom panel of
Fig. \ref{fig:sawtooth_L848}, where we give the
one-dimensional marginalized likelihood distributions 
such as
\begin{equation}
{\cal L}(t_{\rm peak}) \propto \int {\cal L}(t_{\rm peak}, \dtpar, \Phi_{\rm peak}) \ d \dtpar \ d\Phi_{\rm peak} \ \ .
\end{equation}
We see that unlike the well-determined peak time and flux, the sawtooth width distribution shows
a rapid rise and a long tail toward long durations.

The middle panel of Fig. \ref{fig:sawtooth_L848}
shows the data for this core, overlaid with the curve
for the best-fit parameters.  We see that this optimal sawtooth function ``turns on'' 
essentially at the earliest nonzero measurement (largest time before present),
denoted $t_{\rm first}$ in Table \ref{tab:data-timespan}. The sawtooth onset is the peak time $t_{\rm peak}$, and so for the best fit, this essentially is set by the first nonzero measurement.  The best-fit curve goes to zero soon after the last nonzero point (smallest time before present). This means that the width parameter $\dtpar$ is in this case essentially  set by the time interval between the first and last nonzero measurements shown in Table \ref{tab:data-timespan}.  We see that the height of the sawtooth at onset is a compromise among the data points so that, unsurprisingly, the measurements with the smallest errors determine the peak height $\Phi_{\rm peak}$ and also influence the slope and hence the width.

We can understand the broad width $\dtpar$ distribution for this and other sawtooth fits by considering the interplay between the data and the best-fit curve in the middle panel of Fig.~\ref{fig:sawtooth_L848}.
The nonzero measurements exact a ``cost'' in goodness of fit $C$ for models that miss them, with the most extreme case being outright rejection ($C \rightarrow \infty$) of models 
that predict zero flux where counts are nonzero.
Conversely, the points with zero counts impose a cost in $C$ for fits that are nonzero, but this penalty
is less severe, corresponding to allowing for Poisson fluctuations.  This means that sawtooth fits will be highly suppressed if they are narrower than the nonzero count data, but will have some freedom to extend beyond the width of the data, until the available zero-count points suppress fits that are much wider than the data.  This is the trend we see in the width distribution.

These insights from Fig.~\ref{fig:sawtooth_L848} elucidate
also the trends in sawtooth fits to the other sediment cores.
Fig.~\ref{fig:sawtooth_L851} shows sawtooth fits
for Ludwig core 851.  Here again we see that the peak time and peak flux values are fairly well determined; though the top and bottom panels show that the peak time has a sharp lower limit, but its distribution extends for about 0.4 Myr beyond this.  We can understand this feature from the middle panel:  there is a lack of data between the earliest measured nonzero point and prior zero-points. This gap is about 0.3 Myr, and the lack of data here means that there is no  penalty for fits that have an onset $t_{\rm peak}$ anywhere in this range.  This leads to the width in $t_{\rm peak}$.\footnote{
The top panels of Fig.~\ref{fig:sawtooth_L851} also show that the width is positively correlated with peak time.
This reflects the fact that, to maintain a similar shape of curve through the nonzero data, a larger width 
is compensated by an earlier peak time.}

Even more striking is the width parameter in Fig.~\ref{fig:sawtooth_L851}: we see in the bottom panel that the width distribution comes to a peak at 1.84 Myr, but the distribution is highly skewed.
The $\dtpar$ likelihood cuts off rapidly below the peak, but shows a long tail beyond the peak that extends out to the longest values that were allowed.  Again, the data in the middle panel show the reason: after the last nonzero point (earliest time before present), there is only a single point with zero counts.  Thus there is little penalty for fits with a width extending far beyond the interval between the nonzero points.  Furthermore, the scatter of the nonzero points allows for a wide range of slopes, which also permits large widths.
The lesson here is clear and reasonable:
{\em to get a strong constraint on the width is it essential to measure multiple points with zero \fe60 counts before and after the points with nonzero counts.}  For Ludwig core 848, this is the case, and the width is better constrained than that of Ludwig core 851.

This lesson is underscored when we consider Figs.
\ref{fig:sawtooth_W4521} and \ref{fig:sawtooth_W4953},
which show sawtooth fits to Wallner cores 4521 and 4953 respectively.  For both cores, we see that effectively
we can only set a lower limit to the width parameter. That is, the width likelihood in the bottom center panel begins to rise after some minimum $\dtpar$, and continues to increase up to the highest allowed value.  Looking at the data, we see that both cores have no points with zero counts before or after the nonzero counts.  Thus, the nonzero count duration sets a lower limit to the width, corresponding to the onset of the likelihood rise, about $(1.6, 2.5)$ Myr for cores (4521, 4593).  However, the available data essentially set no upper limit to the width for these cores.
We also see that the peak time is poorly constrained, again due to the lack of zero-count data at times earlier than the first nonzero point.

We have produced plots in the style of Figs.~\ref{fig:sawtooth_L848}--\ref{fig:sawtooth_W4953} for the other fitting functions.
For brevity we show here only
the Gaussian fit to Ludwig core 848, which
appears in Fig.~\ref{fig:gaussian_L848}.
The main trends are similar to those we saw for the sawtooth fit to this core in Fig.~\ref{fig:sawtooth_L848}, and
the peak width and peak flux are quite well determined. 
For the Gaussian case, we also find that the width
is well determined, better than the sawtooth case. 
For the cores not shown, it is illuminating to compare the trends in the Gaussian fit to those found in the sawtooth fits shown in Figs.~\ref{fig:sawtooth_L851}--\ref{fig:sawtooth_W4953}.
For the Ludwig core 851, the peak time and width are broader than those in core 848, but are still well determined.
On the other hand, for Wallner core 4521, 
the Gaussian fits also give only a lower limit
to the width, and the peak time likelihood does
have a clear maximum but broad tails on either side.
Interestingly, the Wallner core 4953 fits are 
better determined, with the width and other likelihoods
showing clear peaks and wings that go to zero on both sides.  We believe this is due to the effect of a few nonzero points with high-precision fluxes, which anchor the Gaussian fit and prevent large excursions away from the best-fit region.

Among the other fits we tried, the reverse sawtooth case also deserves mention. Here the fit has a linear rise at early times, ending with an abrupt cutoff at late times; this was chosen to contrast with the more physically  motivated sawtooth case.  
For these fits, we find that the width parameter
likelihoods set only lower limits for all cases.
This includes the Ludwig cores for which
it had been possible to determine the width in the
ordinary sawtooth case.

Figure \ref{fig:all_fits} summarizes the best-fit curves for the six different fitting functions.  Several trends emerge.
Regarding the pulse widths, we see that, for the symmetric functions (Gaussian, Lorentzian, triangle), the best-fit curves all span at least 1 Myr, with Wallner cores spanning $>2 \rm \ Myr$.  For the asymmetric functions (sawtooth, reverse sawtooth, cut exponential), the best-fit curves are generally even wider.
The initial signal arrival is sharply defined in the sawtooth and cut exponential cases, where it exceeds 3 Mya
for at least one Wallner and one Ludwig core.  In the other cases, the onset is gradual but also begins no later than 3 Mya.

Turning to the signal amplitude, we see that in all cases the Wallner peak fluxes are higher than those of Ludwig. These differences were discussed in \S\ref{sec:data_intro}, and may point to geophysical differences in \fe60 fallout, and could also reflect differences in \fe60 extraction techniques. We note that the \fe60/Fe ratios in the two Ludwig sediments are similar, as are the isotope ratios in the two Wallner sediments.  The differences between the {\em fluxes} in the two Ludwig (and two Wallner) cores are thus largely due to the differences in sedimentation rates.  For example, the sedimentation rate for L-848 is a factor $\sim 3$ lower than that of L-851.  These differences propagate to our fits summarized in Fig.~\ref{fig:all_fits}, with the Gaussian and Lorenzian profiles showing similar ratios.  But the fit shape has an influence, as the triangle fits give lower ratios of peak flux, particularly for the Wallner cores, while the cut exponential and sawtooth fits give larger ratios.~\footnote{We are grateful to the referee for pointing this out to us.}

To provide a basis for comparing quantitatively the different fits for each core, we compiled all of the best-fit \cmin values for the six fits shown in Fig.~\ref{fig:cmin} and Table \ref{tab:cmin}; for each data set, $C$ measures the negative of the log of the likelihood, in a close analogy to a $\chi^2$ for continuous data.  The minimum value \cmin thus identifies the point of maximum likelihood, playing the role of the minimum $\chi^2$.  We note that, for a given core, \cmin quantifies goodness of fit, in that the more negative the \cmin value, the better the fit.  As with $\chi^2$, variations of $C-C_{\rm min} \lesssim 1$ around the minimum are not statistically significant, while larger variations indicate a preference for the model with lower \cmin.  We also note that, while the \cmin can be compared between fits for the same core, it is not appropriate to compare the relative \cmin values for fits between different cores (for example, the fact that the \cmin values for the Ludwig cores are above zero and the \cmin values for the Wallner cores are below zero says nothing about the relative goodness of fits to the Wallner and Ludwig cores).  In Table \ref{tab:cmin}, the most negative value for each core has been shown in boldface.  We show in italics \cmin values that are nearly identical for the respective core have been shown in italics, indicating that the respective fit functions for the bold and italic values are all of similar quality.  

We find that, for Wallner Core 4521, the triangle, Gaussian, and Lorentz fits are all equally good. As seen in Fig.~\ref{fig:sawtooth_W4521}, the data for Core 4521 are quite irregular, which could account for the multiple favorite fit shapes.  Wallner Core 4953 also has a preference for the triangle and Gaussian fits.  Meanwhile, the Ludwig Cores 848 and 851, which are more heavily sampled than the Wallner cores, have a clear preference for the sawtooth and Lorentz fits, respectively.  The main takeaway from these results is that the \fe60 pulse does not have a preferred shape, even across the same core.  
Although we frequently describe the widths of the six fitting functions as the width timescales, it should be noted that these are not actually good measures for comparing different fit shapes.  For example, $\dtpar$ for the sawtooth fit is the full width (FW) of the fitting function and therefore is the actual timescale for that curve shape. However, $\dtpar$ for the Gaussian is not given directly by the actual start and stop times of the function, since the Gaussian never reaches zero flux.  Therefore, in order to compare the timescales for each function (and find a preferred time width for the supernova pulse), we have examined the traditional FW at half maximum time for each fit for each core.  Since our primary interest is the maximum width of the function, we have also plotted the FW at 0.1 the maximum (which is closer to the true timescale). Table \ref{tab:fwhm} and Fig.~\ref{fig:fwhm} show differences in the width determination between the Wallner and Ludwig samples. The values corresponding to the best-fit \cmin values from Table \ref{tab:cmin} are also given in boldface and italics where relevant.  As we have noted, these reflect differences in sampling rather than a true discrepancy; hence the robust conclusion is the lower limit this places on the signal duration.   

Thus the presently available sediment data carry enough uncertainty (particularly in sampling) to prevent an unambiguous measurement of the preferred shape or timescale for all cores.  As future measurements are made, these questions will become all the more important, because a demonstrated consistency among the measurements will allow us to probe the underlying astrophysics.

{\it Our key result is that the FW at 0.1 maximum height, for all functions and cores, shows that the width of the deposition timescale is at least 1 Myr.  This is significantly longer than the prediction of the traditional Sedov model.}  

\begin{table}[htbp]     
    \centering
        \caption{\textit{Goodness of Fit:  \cmin Values for 3-parameter fits.} }
    \hspace{-25mm}
    \begin{tabular}{c|cccc}
    \hline \hline 
        &  \multicolumn{4}{c}{\cmin Values} \\
    \hline 
        & \multicolumn{2}{c}{\citet{wallner2016}} & \multicolumn{2}{c}{\citet{ludwig2016}} \\
      Model & Core 4521 & Core 4953 & Core 848 & Core 851 \\
    \hline
      Cut Exponential   & -311.17 & -158.58 & 63.33 & 30.32 \\
      Gaussian  & \textit{-312.56} & \textit{-174.25} & 62.24 & 26.55 \\
      Lorentz   & \textit{-312.45} & -173.15 & 65.04 & \textbf{24.62} \\
      Sawtooth  & -311.02 & -158.26 & \textbf{57.73} & 29.00 \\
      Reverse Saw   & -311.72 & -153.66 & 80.61 & 39.01 \\
      Triangle  & \textbf{-312.59} & \textbf{-174.34} & 59.85 & 25.56 \\
    \hline \hline
    \end{tabular}
    \vspace{3mm}
    \label{tab:cmin} \\
    \parbox{0.9\textwidth}{The  most negative \cmin value for each core gives the {best fit}, which is shown in {\bf boldface}.  Fits that have nearly identical \cmin (for the respective core) are in {\it italics}.  Best-fit values should be compared between models for the same core (vertical columns), and not between cores (horizontal rows).}
\end{table}

\begin{table}[htbp]     
    \centering
        \caption{\textit{Best-fit Timescale Width Values for 3-parameter fits.} }
    \hspace{-25mm}
\begin{tabular}{c|cccc}
    \hline \hline 
        &  \multicolumn{4}{c}{FWHM (Myr)} \\
    \hline 
        & \multicolumn{2}{c}{\citet{wallner2016}} & \multicolumn{2}{c}{\citet{ludwig2016}} \\
      Model & Core 4521 & Core 4953 & Core 848 & Core 851 \\
    \hline
      Cut Exponential   &  2.60 $\pm$ 0.48 & 2.60 $\pm$ 0.33 & 0.42 $\pm$ 0.20 & 0.92 $\pm$ 0.58\\
      Gaussian  &  \textit{1.17 $\pm$ 1.97} & \textit{1.17 $\pm$ 1.52} & 0.73 $\pm$ 0.13 & 0.91 $\pm$ 1.76\\
      Lorentz   &  \textit{1.45 $\pm$ 1.63} & 1.15 $\pm$ 1.24 & 0.47 $\pm$ 0.17 & \textbf{0.55 $\pm$ 1.20} \\
      Sawtooth  & 1.88 $\pm$ 0.24 & 1.88 $\pm$ 0.14 & \textbf{0.68 $\pm$ 0.10} & 0.92 $\pm$ 0.30 \\
      Reverse Saw   &  1.88 $\pm$ 0.27 & 1.88 $\pm$ 0.12 & 1.26 $\pm$ 0.31 & 1.88 $\pm$ 0.24\\
      Triangle  & \textbf{2.44 $\pm$ 0.71} & \textbf{1.02 $\pm$ 0.55} & 0.69 $\pm$ 0.09 & 0.83 $\pm$ 0.57 \\
    \hline
        &  \multicolumn{4}{c}{FW at 0.1 Maximum (Myr)} \\
    \hline 
      Cut Exponential   & 8.63 $\pm$ 1.59 & 8.63 $\pm$ 1.09 & 1.41 $\pm$ 0.65 & 3.04 $\pm$ 1.91 \\
      Gaussian  & \textit{2.14 $\pm$ 3.59} & \textit{2.14 $\pm$ 2.77} & 1.34 $\pm$ 0.23 & 1.66 $\pm$ 3.21\\
      Lorentz   & \textit{4.34 $\pm$ 4.90} & 3.44 $\pm$ 3.71 & 1.42 $\pm$ 0.50 & \textbf{1.65 $\pm$ 3.59}\\
      Sawtooth  & 3.38 $\pm$ 0.43 & 3.38 $\pm$ 0.26 & \textbf{1.22 $\pm$ 0.19} & 1.66 $\pm$ 0.54\\
      Reverse Saw   & 3.38 $\pm$ 0.49 & 3.38 $\pm$ 0.22 & 2.27 $\pm$ 0.56 & 3.38 $\pm$ 0.44 \\
      Triangle  & \textbf{4.40 $\pm$ 1.28} & \textbf{1.84 $\pm$ 0.99} & 1.23 $\pm$ 0.17 & 1.50 $\pm$ 1.03 \\
    \hline \hline
    \end{tabular}
        \vspace{3mm}
\label{tab:fwhm} \\
\parbox{0.9\textwidth}{Values in {\bf boldface} match the best-fit \cmin in Table \ref{tab:cmin}; similarly, values in {\it italics} are nearly identical best fits for that core.  The errors are 1$\sigma$.}
\end{table}

\subsection{Terrestrial and Geophysical Effects---Is the Signal Width of an Astronomical Origin?}
\label{subsec:itaintgeo}

 We now discuss how terrestrial and geophysical effects might smear the timescale of the
  \fe60 pulse.  

\textit{The atmosphere}. When the dust, which is traveling at up to 100 km s$^{-1}$, hits the Earth's atmosphere, it is vaporized.  The iron atoms then combine in the upper atmosphere with molecules such as ozone and hydroxyl radicals, and are eventually deposited on land and in the ocean.  \citet{fry2016} provides excellent approximations for the residence times of the iron in the atmosphere and demonstrates that the iron settles out of the atmosphere in less than 10 yr.

\textit{Deposition on land.} There is no currently known way to detect the \fe60 signal on land, unless it is deposited on an ice sheet, e.g., in Antarctica, where ongoing deposition has recently been measured \citep{koll2019}.  

\textit{Deposition in the ocean.} The residence, i.e., removal, time of iron in the ocean is relatively short, on the order of  $\sim$ 500 years at most \citep{bruland1994,boyle1997,resing2015}. When the dust settles on the ocean floor, it may be taken up by FeMn crusts or deposited as sediment.  

\textit{FeMn Crusts.} Once iron is absorbed by a FeMn crust, it remains ``locked-in'' and unchanged until analysis, and the time signal is accurately preserved.  However, due to the slow growth rate, it is very difficult to time resolve FeMn crusts on the order of kyr, and only the recent work of \citet{wallner2021} has done so.  There is also the possibility of crust porosity, which would enable the iron to attach below the surface, thus smearing the signal.

\textit{Sediments.} It is possible to resolve the time structure of deposits in sediments on the order of kiloyears, but there are a number of effects that can smear the time signature.  The two most important are bioturbation (the churning of the upper few layers of sediment by macroscopic organisms) and chemical reducing environments (created by bacteria and leading to the movement of iron within the sediment column). However, all the sediment samples we consider are from the deep sea where bioturbation effects are negligible, and were carefully selected to ensure that a reducing environment did not occur \citep{fitoussi2008,ludwig2016}.

In summary, there are a number of geochemical, geophysical, oceanic, and biological effects that can in principle distort the \fe60 timescale.  However, the combination of these effects is less than $\Delta t \sim 10^{3}$ yr, which is far shorter than the $\Delta t \sim 10^{6}$ yr timescale found in the data. 
We conclude, therefore, that the measured timescale \textit{must} be astrophysical in origin.
Accordingly, we need a model for the origin and transport to Earth of the dust that can accommodate the signal width and might also be able to predict the line shape, which could be constrained by future data with higher precision.

\subsection{Multiple supernovae?} \label{subsec:multiple_SN}

There is significant interest in analyzing the possibility of multiple supernovae to account for the signal $\sim 3$~Myr ago, e.g., 
\citet{breitschwerdt2016} and
\citet{schulreich2018}
propose that some 16 to 19 supernovae have exploded in the Local Bubble, and have contributed to the extended \fe60 signal. 
Unfortunately, the data are too noisy to cleanly distinguish multiple superposed supernovae peaks.
We have
seen that the sediment measurements cannot
unambiguously distinguish
the relatively simple
pulse shapes we have tried,
which have shapes as different as possible for a singly peaked structure.
Thus, the data in hand cannot exclude
a more complex pulse shape that would
superpose multiple supernova
pulses.

Despite these limitations,
the \fe60 data carries
substantial information
bearing on the question of multiple
supernovae creating the observed
3 Myr pulse.
In particular (a) singly peaked pulse shapes provide
adequate descriptions of the 
measurements, and (b)
none of the data show clear 
groupings of points within the 
time range of detected points.
The available data are thus
consistent with a single peak,
and do not {\em require}
multiple events. 

Furthermore, an accounting
for the \fe60 data 
must explain not only the long signal width
seen in the samples
but also the discovery of distinct pulses at 3 and 7 Myr, with no
apparent signal in between.
If there are multiple
events in the 3 Myr peak, 
then there would likely need to 
be a similar set of events for the 7 Myr peak, but then the gap between the two would need explanation.  
These considerations will inform models for multiple supernovae, 
but do not rule them out.

We do not consider it a productive exercise to fit for multiple supernovae
given the limitations of the available data,
so we restrict our analysis to a single supernova.
We discuss models for the pulse width below in \S\ref{sec:pinball}.

Additional time-resolved measurements of the 3 Myr ago signal (such as could be provided by more dense sampling) could help enormously, but would require significant extra efforts beyond those already made to gather the current data set. Another possibility would be to measure the fluxes of additional isotopes across the \fe60 signal region. If there were significant variations in the isotope ratios, these could constitute evidence for multiple supernovae with different combinations of nucleosynthesis mechanisms.

\subsection{Four-parameter Fit} \label{subsec:4_parm}

In order to examine the ambiguity of the preferred fit shape for the data, we include a 4-parameter ``sharktooth" fit in our analysis.  The purpose of this fit is to analyze whether the data is better described by an asymmetric shape, and at what level of preference.  We chose to work with a 2-width triangle shape, of which the sawtooth, reverse sawtooth, and symmetric triangle are the three-parameter extremes.  The function is defined below in Appendix A.
The shape was given the full range from almost full sawtooth to almost full reverse sawtooth, with only a minor initial minimum of 0.05 Myr to prevent either width from being zero. As with the 3-parameter fits, we also enforced a maximum total width of 3.5 Myr, which is clearly seen to be relevant for Wallner Core 4521 in Fig. \ref{fig:shark_all}.  This width cap is to prevent the fit from stretching all the way to $t=0$ Myr for the noisier data, and was picked as the minimum width needed before the triangle shapes stopped changing dramatically.  

\begin{figure}[htbp] 
	\centering
  \includegraphics[width=18cm]{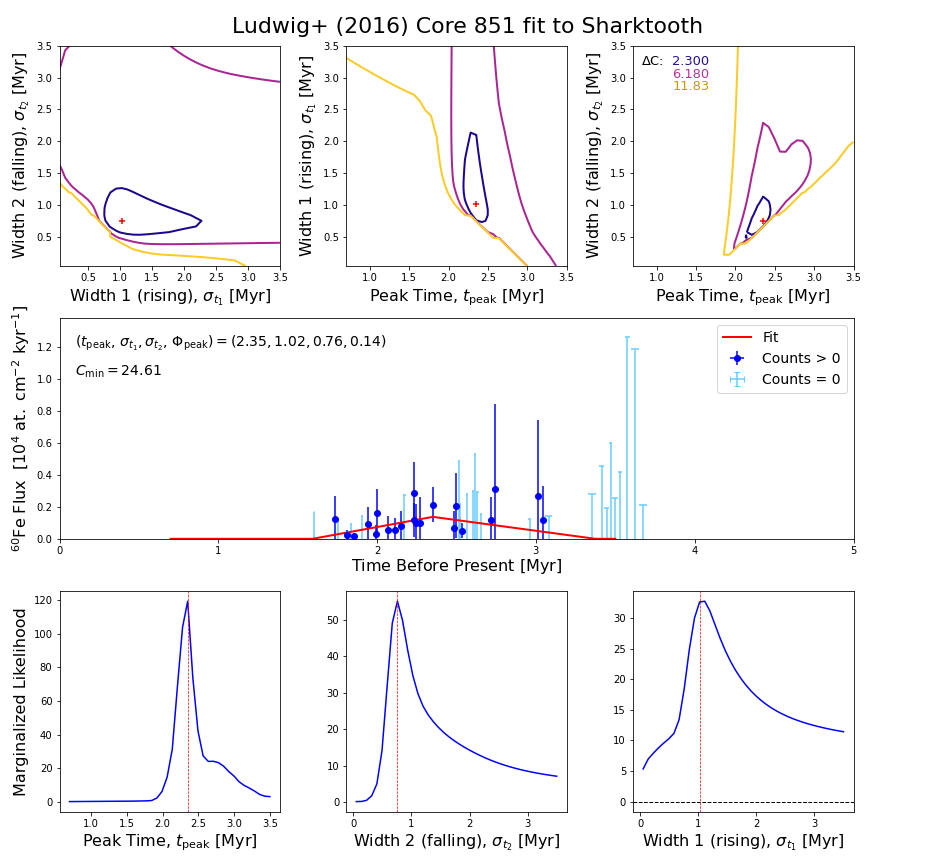} \\
  \caption{\textit{4-parameter sharktooth fits to the terrestrial $^{60}$Fe flux of \citet{ludwig2016}}.  The figure is formatted similarly to the 3-parameter fits, with the two width parameters and the peak time analyzed.  Width 1 ($\sigma_{t_{1}}$) is the rising width for the sharktooth, i.e., the time elapsed between the start of the infall and the peak, and Width 2 ($\sigma_{t_{2}}$) is the time elapsed from the peak until the end of the infall.
  \label{fig:shark}}
\end{figure}

\begin{figure}[htbp] %
	\centering
  \includegraphics[width=16cm]{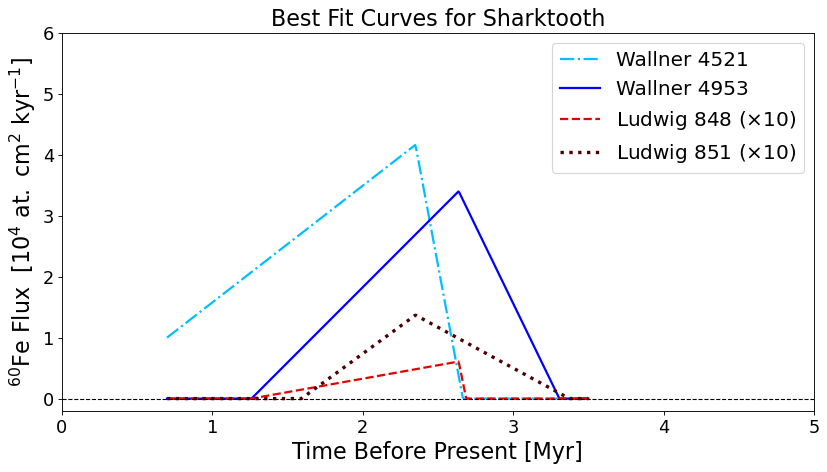} \\
  \caption{\it 4-parameter sharktooth fits to the terrestrial $^{60}$Fe flux.
  \label{fig:shark_all}}
\end{figure}

Results for the Ludwig 851 core appear in Fig.~\ref{fig:shark}.  We see in the bottom panels that the two widths have well-defined peak likelihoods, but broad distributions.  In the upper left panel, we see that the two widths have some anticorrelation, as one might expect if they have to at least sum to the spread in the nonzero data points.  

Examining the best-fit shapes for the four cores in Fig. \ref{fig:shark_all}, we can see that most of the data are best fit with a sharktooth shape that is intermediate between being fully symmetric or asymmetric.  Only Ludwig Core 848 prefers a perfect sharp sawtooth shape, with the fit function capped by the minimum width initial parameter.  The two Wallner cores prefer a slightly sawtooth shape, while Ludwig Core 851 actually has a slight preference toward a reverse sawtooth.    

It should be noted once again that the sharktooth fits are not statistically comparable to the 3-parameter fits in \S \ref{subsec:fit_results}. However, in view of the lack of strong preference for a specific 3-parameter fit, it is an interesting possibility to explore further.  Given the variations in the current data, we cannot meaningfully pick a preferred shape for the \fe60 pulse, so the pulse shape does not provide significant information on the underlying astrophysics.  However, the different pulse widths extracted using the different preferred fit shapes can be used to make some inferences on the astrophysical processes inside the supernova remnant.

\subsection{Global Fits}

\label{sect:global}

\begin{figure}
    \centering
    \includegraphics[width=0.8\textwidth]{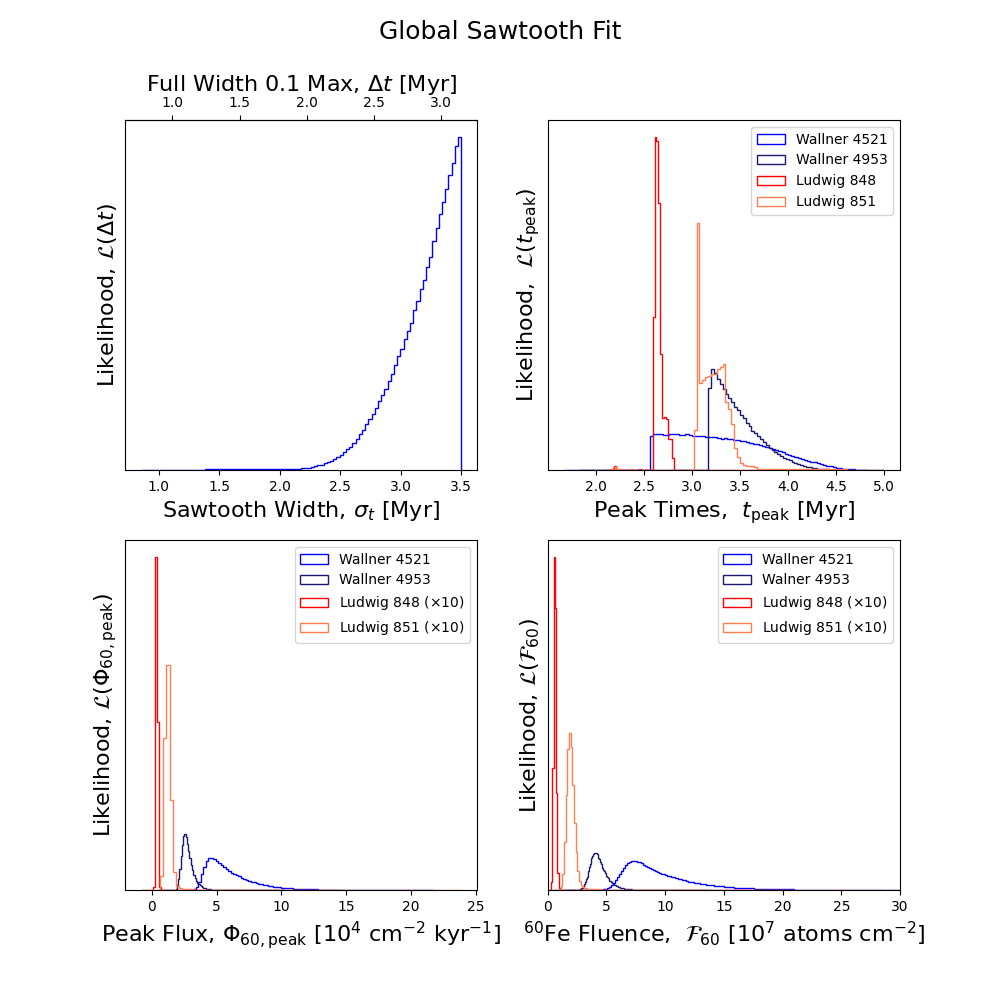}
    \caption{ {\it Global fit to all well-sampled sediments, using
    a sawtooth profile for the flux.} The width is fixed, but the peak flux and peak time are both allowed to vary, allowing for nonuniform global dust fallout and systematic errors in absolute timing, respectively. 
    }
    \label{fig:FixW_VarPH_sawtooth}
\end{figure}

Having examined the cores individually,
we now turn to global fits that use all the cores
to analyze jointly the \fe60 deposition history.
To do this, we first note that the \fe60 flux time profile (i.e., its shape) should be common to all of the sediments regardless of their location 
on the Earth.  
We therefore fit all of the samples
using a single pulse shape 
and thus the same width parameter $\dtpar$.

\begin{table}[htbp]
    \centering
    \caption{Global Fit Parameters}
    \begin{tabular}{cc|ccc}
    \hline \hline
& & \multicolumn{3}{c}{Fit Type} \\
\multicolumn{2}{c|}{Parameter} & Gaussian & Sawtooth & Triangle \\
\hline \hline
Width parameter $\dtpar$ & [Myr] & $0.47 \pm 0.05$ & $3.18 \pm 0.26$ & $0.99 \pm 0.10$ \\
Full width at 0.1 max $\Delta t$ & [Myr] & $2.00 \pm 0.21$ & $2.86 \pm 0.23$ & $1.8 \pm 0.2$  \\
$\Delta t$ 95\% CL lower limit & [Myr] & $1.7$ & $2.7$ & $1.6$ \\
 \hline
 Peak time $t_{\rm peak}$ [Myr] & L-848 & $2.16 \pm 0.10$ & $2.7 \pm 0.05$  & $2.1 \pm 0.09$ \\
   & L-851 & $2.44 \pm 0.11$ & $ 3.2\pm 0.2$  &  $2.4 \pm 0.9$ \\
   & W-4521 & $2.12 \pm 0.11$ & $3.3 \pm 0.5$  & $2.2 \pm 0.9$ \\   
   & W-4953 & $2.44 \pm 0.05$ & $3.5 \pm 0.2$  & $2.5 \pm 0.4$ \\
   \hline
 Peak flux $\Phi_{\rm peak}$ [atoms cm$^{-2}$ kyr$^{-1}$] & L-848 & $0.037 \pm 0.006$ & $0.039 \pm 0.021$ & $0.042 \pm 0.007$  \\
  & L-851 & $0.116 \pm 0.019$ & $0.13 \pm 0.04$ & $0.12 \pm 0.02$  \\
   & W-4521 & $3.89 \pm 0.41$ & $6.0 \pm 1.9$ & $4.5 \pm 0.5$  \\
   & W-4953 & $3.19 \pm 0.42$ & $2.7 \pm 0.5$ & $3.6 \pm 0.5$  \\
   \hline \hline
 Fluence ${\cal F}_{60}$ [$\rm 10^7 \ atoms cm^{-2}$] & L-848 & $0.043 \pm 0.07$ & $0.060 \pm 0.023$ & $0.041 \pm 0.007$  \\
   & L-851 & $0.14 \pm 0.03$ & $0.19 \pm 0.05$ & $0.12 \pm 0.02$  \\
   & W-4521 & $4.4 \pm 0.6$ & $9.5 \pm 3.2$ & $4.4 \pm 0.6$  \\
   & W-4953 & $3.7 \pm 0.4$ & $4.4 \pm 0.8$ & $3.6 \pm 0.3$  \\
         \hline \hline
Goodness of fit  & \cmin  & {\em -195.5} & -185.6 & {\bf -196.8} \\
\hline \hline         
    \end{tabular}
        \vspace{3mm}
    \label{tab:globalfits} \\
    \parbox{0.9\textwidth}{ {\it Fits with a single width parameter but varying peak times and peak flux values for different cores.}  L denotes \citet{ludwig2016}, W denotes \citet{wallner2016}. 
    We list below the best-fit values and 1$\sigma$ uncertainties.
    The peak flux is given in units of $10^4$ atoms cm$^{-2}$ kyr$^{-1}$, and the fluence in units of $10^7$ atoms cm$^{-2}$.
    For the goodness of fit \cmin, {\bf bold} indicates the highest likelihood, and {\em italics} the close second.
}
\end{table}

We account for potential systematic differences between the samples
by allowing the other fit parameters to vary
independently for each core. 
To include  possible systematic offsets in the absolute dating
of the different samples, we fit different peak times for each core; this would
correspond to shifts in the inferred time history between the
samples. Small differences in the peak times of individual fits 
would suggest that differences in the dating are small,
but large differences would point to the presences of 
unaccounted systematics, or the use of a bad fitting function.
We also allow for differences in the peak heights, which could
arise due to different infall and uptake
at different locations.
To perform these fits, we use 
a Markov Chain Monte Carlo approach,
which enables us to search efficiently the 9-dimensional
parameter space using the {\tt emcee} package
\citep{foreman-mackey2013}.\footnote{\href{https://emcee.readthedocs.io}{https://emcee.readthedocs.io}}

Fig. \ref{fig:FixW_VarPH_sawtooth} shows our results for the case of 
a sawtooth time profile; the best-fit parameters and other statistics
are given in Table \ref{tab:globalfits}.  
Turning first to the width, we see that the likelihood is zero until about 2 Myr, then
rises until the highest allowed value.  Thus, as we have seen with the individual fits
in Figs.~\ref{fig:sawtooth_L848}--\ref{fig:sawtooth_W4953},
the global sawtooth form only sets a lower limit to the signal width.  
We find that the 95\% confidence lower limit to the FW at 0.1 maximum
is 2.7 Myr.  This limit is significantly larger than the time span of nonzero data points
shown in Table \ref{tab:data-timespan}, showing that the requirement of a sawtooth form 
leads to wider deposition time.

The shapes of the peak times 
are similar to those in the individual core fits in Figs.~\ref{fig:sawtooth_L848}--\ref{fig:sawtooth_W4953},
and show the same asymmetries.
We find that the most probable peak times differ significantly among the samples.  This reflects the fact that, for a sawtooth, the peak is also the 
point with the earliest nonzero \fe60 counts, itself an artifact of the sampling.
These differences between the peak times (upper right panel of Fig.~\ref{fig:FixW_VarPH_sawtooth})
are also similar to those in the individual fits.

The
peak \fe60 flux values (lower left panel) and the \fe60 fluences
(lower right panel) measured in the two Wallner samples are similar,
and also those of the two Ludwig samples.
However, we do see significant differences in the
peak \fe60 flux values (lower left panel) and the \fe60 fluences
(lower right panel) between the Wallner and Ludwig measurements.
As noted above, this may be due to differences in the analysis techniques, and also
perhaps due to differences in the deposition rates caused by varying
infall or uptake factors. 
We note that the global sawtooth fits for the Wallner cores 
give values that are slightly higher than in the individual fits, but well within errors.

We have also performed global fits for
the Gaussian and triangle fit
functions.  
The results are summarized in
Table \ref{tab:globalfits},
and the Gaussian fit is shown as Fig.~\ref{fig:FixW_VarPH_gauss}.
For the Gaussian case, we see that the FW at 0.1 maximum of $2.00 \pm 0.21 \ \rm Myr$ (upper left panel)
is a compromise between the values of the Gaussian widths found in
individual sample fits as seen in Table~\ref{tab:fwhm}.
On the other hand, the triangle case gave a well-determined width, similar to the global Gaussian fit and the individual triangle fits.
We find that, for the Gaussian fit, the 95\% confidence-level lower limit 
to the FW at 0.1 maximum is 1.7 Myr,
while for the triangle fit the same limit is 1.6 Myr.
The upshot is that, for these other functional forms, the global fits once again give a long timescale for the \fe60 deposition.
Thus, using all of the data, we find that the pulse
timescale is at least 
\begin{equation}
    \Delta t > 1.6 \ \rm Myr \, ,
\label{eq:dtmin}
\end{equation}
which is also close to the interval between
the earliest and latest nonzero \fe60 counts across
all crusts shown in Table \ref{tab:data-timespan}.\footnote{The result in eq.~(\ref{eq:dtmin}) is the full width at 0.1 maximum, and so is slightly less than the  1.65 Myr  global lower limit in Table \ref{tab:data-timespan}. However, the 95\% confidence lower limit on the full triangle width $2\sigma_t > 1.7$ Myr is indeed above this limit.}
Any model for \fe60 delivery must account for this timescale.

For the Gaussian and triangle global fits, we also find that the peak time and peak flux values for each sediment are similar to those
for the corresponding single-sediment fits seen
in Figs.~\ref{fig:sawtooth_L848} to \ref{fig:gaussian_L848}.
The differences among the peak times
are relatively small, with significant overlaps between the different fits. This suggests that the absolute dating of the various samples does not suffer from significant unknown systematic uncertainties.  This also supports our conclusion that the differences in peak times in the sawtooth case (Fig.~\ref{fig:FixW_VarPH_sawtooth}) are due to the abrupt onset of that fitting form, and the sampling.  

Turning to the fluence, Table \ref{tab:globalfits} shows that the Gaussian and triangle results are very similar, indicating that the integral nature of the fluence is not
very sensitive to the differences between these fitting functions.  On the other hand, the sawtooth case gives substantially higher fluences for all sediments.  Since the sawtooth fluence is given by $\Phi_{\rm peak} \, \dtpar/2$, and the peak flux values are consistent across different fitting function, it is clear that the timescale $\dtpar$ is the source of the discrepancy.  Indeed, we have seen that the sawtooth timescale
is poorly determined by individual fits in Figs.~\ref{fig:sawtooth_L848}--\ref{fig:sawtooth_W4953}.  
For this reason, we see that determinations of the fluence are model-dependent given the current data.

In sum, we see that when all of the
sediment data are combined;
the global fits find that the \fe60 fallout time
is long, $\approx 2 \ \rm Myr$.
We recall that, as discussed in \S\ref{subsec:itaintgeo},
 this duration is not a geophysical artifact, but reflects the underlying astrophysical fallout time.  We now turn to the interpretation of
this result.

\section{Implications:  Supernova Dust Formation and Propagation} \label{sec:pinball}

The timescale for \fe60 deposition encodes
information about the delivery of supernova
ejecta from the explosion to its final arrival at Earth. Moreover,
we recall that for supernova material to reach Earth
it must take the form of dust grains
\citep{athanassiadou2011}.  This means
the \fe60 fallout timescale is a probe of the
propagation of supernova dust
over space and time.

We consider in this section two scenarios for supernova dust formation and evolution.
(1) The first scenario adopts the conventional assumption, often implicit,
that supernova dust is entrained in the gas and thus
is comoving with the blast \citep[e.g.,][make this assumption]{fry2015,breitschwerdt2016}.
(2) The other scenario is the model of \citet{fry2020} in which
the dust decouples from the gas, and the trajectories of the charged
grains are determined by the magnetic structure of the remnant.
We develop predictions for the time history
of the dust flux in these two models, which can then be compared with the measured
\fe60 time profiles. 

\subsection{Dust Entrainment Model}

In this picture, the dust grains move with the gas,
so that the grain velocity is the same in magnitude and direction
as the plasma bulk velocity $v_{\rm gr} = v_{\rm gas}$, i.e., the motion is radial, 
overlaid with perturbations due to turbulence.
In addition, we assume the mass density $\rho_{\rm dust}$ of dust grains
is always proportional to the gas density $\rho_{\rm gas} \approx m_p n_{\rm gas}$. 
This model thus posits a direct proportionality
between the mass fluxes of the grain particles and the gas: $J_{\rm gr} = \rho_{\rm dust} v_{\rm gr} \propto \rho_{\rm gas} v_{\rm gas}$.  Thus, the time history of the grain flux is determined by that of the gas flux.

This model is the one adopted by \citet{fry2015},
whose key results we summarize here.
The model focuses on the Sedov phase of the remnant,
in which the blast radius evolves as follows as a function of time $t$:
$r_{\rm blast} = \beta (Et^2/\rho)^{1/5}$,
where $E$ is the kinetic energy of the ejecta into a uniform-density
medium with $\rho = \bar{m} n$, and $\beta = 1.1517$ for monatomic gas, with adiabatic index $\gamma = 5/3$.
Inverting this relation, we derive the following estimate of the blast arrival time at radius $r$:
\beq
\label{entrainment}
t_{\rm blast} = \beta^{5/2} \sqrt{\frac{\rho r^5}{E}}
= 0.036 \ {\rm Myr}
\ \pfrac{\bar{m}}{m_p}^{1/2}
\ \pfrac{n}{0.01 \ \rm cm^{-3}}^{1/2}
\ \pfrac{10^{51} \ \rm erg}{E}^{1/2}
\ \pfrac{r}{50 \ \rm pc}^{5/2} \, ,
\eeq
where we scale using benchmark estimates of the Local Bubble density and total blast energy \citep[see, e.g.,][]{fry2020}.
The corresponding speed of expansion is
\beq
v_{\rm blast} = \frac{2}{5} \frac{r}{t}
= \frac{2}{5}  \sqrt{ \beta \frac{E}{\rho r^3} }
= 550 \ {\rm km/s}
\ \pfrac{m_p}{\bar{m}}^{1/2}
\ \pfrac{0.01 \ \rm cm^{-3}}{n}^{1/2}
\ \pfrac{E}{10^{51} \ \rm erg}^{1/2}
\ \pfrac{50 \ \rm pc}{r}^{3/2} \, .
\eeq
Just behind the shock, the gas density is always the same
constant multiple of the ISM density, namely $\rho_{\rm gas} = 4 \rho_{\rm ism}$
for $\gamma=5/3$.
However, because the mass of the supernova ejecta is fixed initially, the ejecta density
must drop as the blast volume grows.  To fix numbers, 
we assume that the ejecta are entrained with the gas,
and approximate the blast as a thin uniform-density shell
of fractional width $x = \Delta r_{\rm shell}/r_{\rm blast} \ll 1$,
in which case mass conservation implies $x \approx 1/12$.

The duration of the flux is the timescale for the blast shell to
pass by.
At a fixed distance $r$, and using self-similarity,
the shell crossing time is
\beqar
\Delta t_{\rm shell} & 
\approx & \frac{x r}{v(r)} = \frac{5 x}{2} t \approx \frac{5}{24} t \\
& = & 0.03 \ {\rm Myr}
\ \pfrac{m_p \times 0.01 \ \rm cm^{-3}}{\rho}^{1/2}
\ \pfrac{E}{10^{51} \ \rm erg}^{1/2}
\ \pfrac{100 \ \rm pc}{r}^{5/2} \, .
\eeqar
This is nearly two orders of magnitude smaller than the observed \fe60
pulse width.

The ISM density would have to be $\gtrsim 100 \ \rm cm^{-3}$ in order to
overcome this discrepancy.
Such a density is characteristic of a giant molecular cloud complex,
which could have been a plausible density if our \fe60-depositing supernova
were the first to explode in a nearby star-forming region.
However, modeling of the Local Bubble indicates that it has hosted multiple supernovae
over timescales of 3 Myr or longer \citep{smith2001,breitschwerdt2016},
and there is now clear evidence for \fe60 deposition by an explosion $\sim 7$~Mya~\citep{wallner2021}. 
After the first explosion, the local interstellar region would have a much smaller
density, so we are driven to consider other explanations
for the long \fe60 timescale.

It is also worth highlighting that terrestrial (anthropogenic) explosions do not exhibit efficient mixing of ejecta with the larger blast wave.  
Although both move outward rapidly, the ejecta (i.e., the material responsible for the explosion)
remains confined relatively near the center of the explosion while the forward shock
travels much greater distances and without carrying ejecta material with it.  For example, in the case of the 
Chelyabinsk bolide, the larger meteorite fragments were found in a 32-km long, 10 km wide region along the original trajectory
of the meteor \citep{popova2013}.  Meanwhile, the smaller aerosols and dust particles formed a cloud that rose vertically 11 km over 80 s then 
stabilized at that altitude for the next 120 s \citep{gorkavyi2013}.
In contrast, the shock wave produced by the bolide propagated radially outward traveling 23 km in 76 s and 52 km in 173 s \citep{popova2013}
showing a definitive decoupling between the shock wave and ejected material.
Such observations of decoupling further constrain the entrainment model's validity.

Several possible explanations of the long \fe60 deposition timescale have been discussed elsewhere.
(1) The prolongation of the signal over
time could reflect multiple supernovae, each with a narrow pulse width \citep{breitschwerdt2016}. As we have noted,
the data does not demand this, but
also cannot exclude multiple supernovae in the 3 Myr peak. 
However, because we observe a broad \fe60
pulse $\sim 3$~Mya, preceded by a gap and and another pulse $\sim 7$~Mya 
that also seems broader than suggested by the entrainment model (\ref{entrainment}), this scenario would
require two bursts of supernovae
in rapid sequence, with a well-defined lull in between.
(2) The \fe60 flux could
reflect the motion of the Sun through the supernova material \citep{chaikin2022},
with a complex time history needed to accommodate the two broad pulses.
(3) 
Another possibility is that some \fe60-bearing dust was trapped by the Local Interstellar Cloud,
an $\sim 5$ pc feature that envelops the solar system \citep{koll2019,linsky2019}.
One can quantify this by considering the stopping power of the cloud.
For dust grains the size $a_{\rm gr} \gtrsim 0.3 \, \rm \mu m$ needed to overcome
solar radiation pressure, the stopping distance 
due to drag is $60 \ \rm pc$
for $n_{\rm LIC}=0.2 \ {\rm cm}^{-3}$.
Thus we do not expect drag to efficiently stop the grains unless they are much smaller.
(4)  \citet{opher2022} model the effects of a neutral cloud (seeded with \fe60) passing through the solar system, and show that, if the cloud density is very high, it can compress the heliosphere within 1 au. If the cloud is also large enough, the passage can last for the required time.

We consider these scenarios to be worthy of further investigation,
but also not without their challenges.
Here we propose another solution motivated by our recent work~\citep{fry2020},
assess its merits and drawbacks, and offer observational tests.

\subsection{Charged Dust Model}

We propose that the \fe60 arrives at the Earth (and Moon) as part of 
charged dust grains that were created in the supernova, whose
propagation was largely determined by the magnetic structure of
the remnant impact into the surrounding medium.
\citet{fry2020} performed detailed calculations of
the propagation of charged dust in a supernova remnant,
motivated by the \fe60 data.
Here we summarize
the rich physics influencing dust grain evolution
and propagation.  

Dust formation in supernova remnants is a subject of intense ongoing
research, but it is clear that a substantial amount of dust is formed very soon after the explosion, e.g., SN~1987A shows infrared emission consistent with all of the supernova-produced iron being locked into
grains within tens of years after the explosion \citep{matsuura2011, matsuura2017, matsuura2019},
while recently \citet{Niculescu-Duvaz2022} examine a large sample of supernovae
finding that, after $\sim 30$ yr, on average $0.24^{+0.09}_{-0.05} \, M_\odot$ of material is condensed
into dust. 
We therefore follow \citet{fry2020} in assuming that dust is present,
likely with a range of grain sizes, and initially entrained with
the gas from which it formed.  Thus a range of dust compositions, sizes, and velocities is present.

After the dust is created, it 
suffers collisions with gas particles that lead to
drag, sputtering, and charging. However, these effects are minimal
because the dust is comoving with the gas.  But as the supernova remnant evolves,
a reverse shock propagates inward.  This shocks and slows the gas,
thereby decoupling the dust from the gas. 
Grains suffer some damage at the shock, but those
large enough to survive crossing the reverse shocks will then
be subjected to increased drag, sputtering, and charging.  

Given the negligible magnetic field in the inner portion of the supernova remnant,
the dust still travels radially.  However, when the dust grains subsequently encounter 
the magnetized ISM, it acts as a mirror and reflects the dust.
The dust grains then pass back through the remnant until
they encounter the ISM material once more, and are again
reflected by its magnetic field (at each ISM encounter, there is also some probability that the grains become trapped).
The resulting dynamics is that the dust
is confined to the ejecta region, with repeated bouncing
motion, ``pinball'' style, in the ejecta interior.
As they propagate, the dust particles are slowed by drag and
are sputtered, becoming smaller and losing mass.

An order of magnitude calculation illustrates the key features of dust evolution found
in the \citet{fry2020} simulations.  
We model a dust grain as a sphere of radius $a$,
density $\rho_{\rm dust}$, and mean atomic mass $m_{\rm gr}$.
As the dust particle moves through the gas,
it suffers collisions at a rate
\beq
\Gamma_{\rm coll} = n_{\rm gas} \sigma v_{\rm rel}
\approx  \pi a^2 n_{\rm gas} v_{\rm dust} \, ,
\eeq
where in the second expression we approximate the collision cross section with
the geometric cross section,
and assume that the dust is moving much faster than the
gas, 
so that the relative speed $v_{\rm rel} \approx v_{\rm dust} \gg v_{\rm gas} \sim \sqrt{kT/m_p}$; 
moreover, we assume that the gas is dominated by hydrogen.

Drag has collisional and Coulomb components. 
In the assumed limit of fast dust particles 
with $m_{\rm p} v_{\rm gr}^2 \gg kT$
the collisional term is
\beq
F_{\rm coll} \simeq \Gamma_{\rm coll} m_p v_{\rm rel}
\simeq \Gamma_{\rm coll} m_p v_{\rm gr} \, ,
\eeq
where $\vec{v}_{\rm rel} = \vec{v}_{\rm gr}-\vec{v}_{\rm gas}$
is the grain speed relative to the local gas speed.
This gives a collisional stopping time of order
\beqar
\tau_{\rm coll} & =  &\left| \frac{M_{\rm gr} v_{\rm gr}}{F_{\rm coll}} \right|
\simeq \frac{4 \rho_{\rm gr} a}{3 m_{\rm p} n_{\rm gas} v_{\rm gr}} \\
& = & 3.9 \ {\rm Myr} \ \pfrac{a}{0.3 \ \mu \rm m}
\ \pfrac{10^{-2} \ \rm cm^{-3}}{n_{\rm gas}}
\ \pfrac{300 \ \rm km/s}{v_{\rm gr}} \, ,
\eeqar
where $M_{\rm gr}$ is the grain mass.
The Coulomb drag force for high speeds larger than
the plasma thermal speeds $v^2 \gg kT/m_p$
is independent of grain size: $F_{\rm Coul} \approx 4\pi z_{\rm gr}^2 \, e^2 \, n_{\rm gas} \, \ln[\Lambda]/m_p v^2$, in contrast to the
collisional drag; here $\ln [\Lambda] \sim 20$ is the so-called Coulomb logarithm.  The Coulomb stopping time therefore
exhibits a different and stronger dependence on grain size:
\beqar
\tau_{\rm Coul} & =  &\left| \frac{M_{\rm gr} v_{\rm gr}}{F_{\rm Coul}} \right|
\simeq \frac{\rho_{\rm gr} a^3 v_{\rm gr}^3}{3 z_{\rm gr}^2 e^2 n_{\rm gas}} \\
& = & 2.5 \ {\rm Myr} \ \pfrac{a}{0.3 \ \mu \rm m}^3
\ \pfrac{10^{-2} \ \rm cm^{-3}}{n_{\rm gas}}
\ \pfrac{300 \ \rm km/s}{v_{\rm gr}}
\ \pfrac{175 \ \rm V}{U_{\rm gr}} \, .
\eeqar
If the grain
is indeed moving much faster than the gas,
then sputtering decreases the grain radius
at a rate
\beq
\frac{da}{dt} \simeq - \frac{m_{\rm gr} Y}{4\pi \rho_{\rm gr} a^2} \Gamma_{\rm coll}
= - \frac{m_{\rm gr} n_{\rm gas}}{4\rho_{\rm gr}} Y v_{\rm dust} \, ,
\eeq
where $Y$ is the energy- or velocity-averaged
yield of atoms liberated per collision,
and $m_{\rm gr}$ is the mean mass of a grain {\em atom}.
The resulting timescale for grain erosion due to sputtering is
\beq
\tau_{\rm sput} = \left| \frac{a}{\dot{a}} \right|
\simeq \frac{3}{Y} \frac{m_{\rm p}}{m_{\rm gr}} \ \tau_{\rm drag}
= 2 \pfrac{50}{m_{\rm gr}/m_{\rm p}} \frac{0.03}{Y} \ \tau_{\rm drag} \, .
\eeq
We see that, in the high-velocity regime, the 
sputtering and drag timescales are related
by a factor that depends only on the grain atomic mass
and yields.  Estimating these values for iron-bearing grains
shows the
timescales to be comparable, with drag somewhat faster.
We thus expect significant drag and sputtering both to occur,
and that the grain lifetime is similar to the drag timescale.

\begin{figure}
    \centering
    \includegraphics[width=0.32\textwidth]{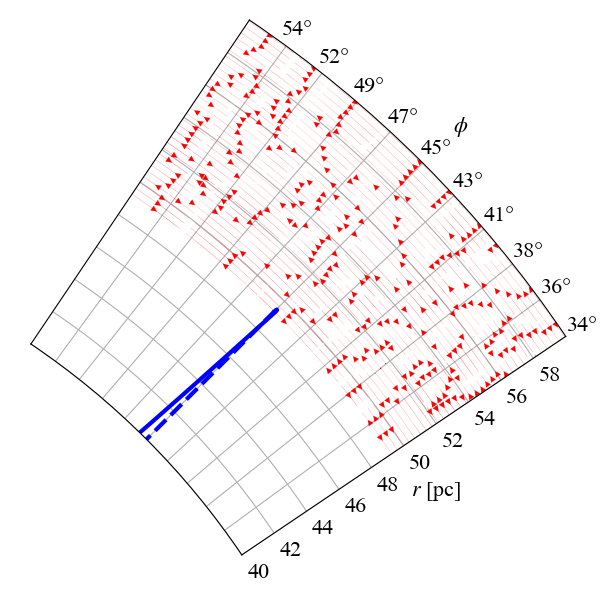}
    \includegraphics[width=0.32\textwidth]{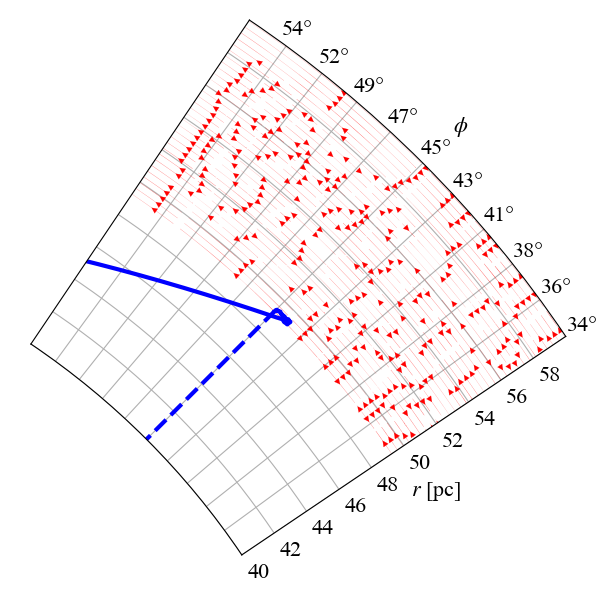}
    \includegraphics[width=0.32\textwidth]{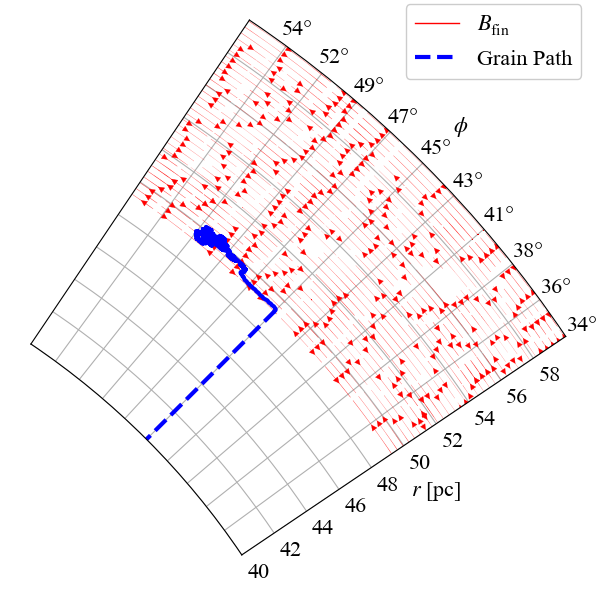}
    \caption{ {\it Three sample azimuthal trajectories from our 0.1 $\mu$m metallic Fe dust grain simulations.} The left panel shows a grain 
    being bounced (reflected off) the magnetic field in the ISM, and the center panel shows a grain becoming temporarily
    trapped by the ISM magnetic field before eventually escaping. Finally, the right panel shows a grain becoming 
    trapped by the ISM magnetic field for the duration of the simulation.  The grain path is shown as a dashed blue line until reflection, and a solid blue line afterwards.  The red lines are the magnetic field lines at the moment of reflection ($t \approx 130$ kyr, $r \approx 50$ pc after the supernova).\\
}
\label{fig:dustpaths}
\end{figure}

Figure \ref{fig:dustpaths} shows sample simulations of the trajectories of 0.1 $\mu$m dust grains (blue) 
encountering the magnetic field in the ISM (red). The left panel shows an example where the dust grain 
bounces (is reflected) straight back from the ISM, and the central panel shows an example where the dust grain
is trapped temporarily by the ISM magnetic field, but eventually escapes. These exemplify the 
``pinball" feature discussed in \citet{fry2020}, whereby the directions of the dust grains' trajectories
can be changed, losing memory of the location of their progenitor. Finally, the third panel shows an example where the
dust grain is trapped for the duration of the simulation. All three panels illustrate how the timescale for the 
deposition of live radioisotopes can be extended due to interactions with the ISM.

Determining quantitatively the conditions when a grain will be reflected versus trapped (and for how long) is a subject 
requiring further examination.  A grain's charge depends on its composition, speed, and the surrounding environment; the 
dynamics of the magnetic field is determined by the supernova remnant's plasma dynamics.  While the grain's possible motions are
understood (i.e., curvature drift, gradient drift, reflections, etc.), when those occur is not fully characterized.  The time
the grain first encounters the magnetic field, the pitch angle of the grain's velocity with the local magnetic field,
the dynamical timescale of the magnetic field, and the size of any turbulent eddies and density perturbations can influence the 
confinement of the dust grains.  \citet{fry2020} provided an initial statistical result for metallic iron grains,
but a more detailed examination is beyond the scope of this work.

To summarize, our model builds upon the results of \citet{fry2020} 
to predict the following supernova dust grain history and dynamics as the supernova remnant evolves.
\begin{itemize}

    \item 
    {\em Free expansion phase: coupled.} Dust grains are nucleated in dense ejecta knots, comoving with gas \citep{fry2020}.
    
    \item
    {\em Start of Sedov phase: decoupling.} As the ejecta (composed of ejected gas and dust grains) experiences the reverse shock and decelerates, the surviving dust grains are dynamically decoupled from the gas.  Their subsequent propagation is dominated by the Lorentz force and drag.  In weakly or nonmagnetized supernova material, their motion is radial until they encounter the magnetized ISM \citep{fry2020}.
    
    \item
    {\em Sedov and snowplow phases:  reflection and trapping.} In encounters with the 
    magnetized ISM, grains may be either reflected or trapped.
    The reflected grains traverse the inner supernova remnant until their next encounter with the ISM.  
    Grain size plays a crucial role here:
    the smallest grains ($a \sim 0.005 \ \mu \rm m$) are trapped at the first encounter, sputtered, and
    lost, whereas the largest grains ($a \sim 1 \ \mu \rm m$) travel
    past the supernova-ISM boundary and then are trapped.
    Intermediate-sized grains ($a \sim 0.05-0.1 \ \mu \rm m$) have non-negligible
    probabilities both to be trapped and to escape.
    Thus over time the grain number density in the supernova material will
    drop, in favor of a buildup of trapped grains at the supernova-ISM boundaries.
    The trapped dust motion depends on the grain size
    and potential (i.e., the charge-to-mass ratio).
   
    \item
    {\em Fadeaway phase: shell buildup and release to ISM.}  Trapped grain
    motion in turbulent field lines will be approximately diffusive,
    leading to a buildup in a shell at the supernova-ISM boundary.
    Spatial and time changes in the magnetic fields
    can lead to grain deceleration and acceleration,
    in addition to the action of drag. 
    The grains are stopped over the drag timescale. The larger grains survive 
    and remain as the forward shock slows to a sound wave, and the supernova remnant fades.
   
\end{itemize}

In this picture, the spatial distribution of dust grain changes over time.  While the dust grains are predominantly reflected, the grains should roughly uniformly populate the inner supernova remnant. 
As they become trapped in the surrounding magnetized ISM, the grains should move diffusively in a shell of increasing thickness around the inner remnant. 
Thus, when the grain-bearing material arrives at the solar system, we expect the particle density to be a mix of a uniform and shell profile, with the shell profile more favored at late times and large distances.
Detailed modeling of this distribution,
and the resulting \fe60 time profile at Earth,
is beyond the scope of this paper, but is a problem we intend to revisit in future work.

\section{Consequences and Tests} \label{sec:tests}

The sequence of events in the pinball model for supernova grain evolution
has significant implications for different dust species
and radioisotope signatures, as we now discuss.

\textit{Supernova distance.}  Rough estimates have suggested that the origin of the spike in \fe60
$\sim 3$ Mya might have been a supernova that exploded
within about 100 pc of Earth. Several stellar clusters are known to have
passed within 200 pc of Earth within the past 35 Myr. Two of these have 
attracted particular attention: the Tuc-Hor group
\citep{mamajek2015,hyde2018}, which was within
$\sim 60$ pc of Earth at the time of the event that produced the 3 Mya signal, and the Sco-Cen OB association \citep{benitez2002}, which was 
$\sim 130$ pc away at that time --- the possibility of a runaway 
star has also been considered. No conclusive evidence in favor of
any hypothesis has been found.

Our analysis of the propagation of magnetic dust indicates that it
could not progress far into the ISM. See, in particular, Fig.~4 of
\citet{fry2020}, where simulations of metallic Fe grains of varying 
initial sizes indicate that they would reach a maximum distance of
about 50 pc. This limited range favors a Tuc-Hor origin over the
Sco-Cen hypothesis, while being consistent with the runaway star hypothesis.

\textit{Gamma-ray line spectroscopy of nuclear lines.}
There have been multiple observations of gamma rays from decays
of \fe60 and \al26 in the ISM. In our model, the dust bearing
\fe60 and \al26 dust moves at high speeds within the supernova 
remnant for much of the radioactive lifetime of the species.  
Therefore we would expect the \fe60 and \al26 nuclear lines to exhibit
Doppler broadening. Indeed, International Gamma-Ray Astrophysics Laboratory (INTEGRAL) data indicate that the
line width of \al26 decay is broadened
\citep{Kretschmer2013}.
If the dust drag time reflects the \fe60 width,
then our work shows that the stopping timescale $\gtrsim 1 \ \rm Myr$.  This is longer than the 0.717 Myr half-life of \al26, which
means that most of the \al26 in supernova dust
will decay while moving at high speed,
consistent with the INTEGRAL result.

\textit{Supernova remnants.}
If supernova dust grains are only created at early times, 
then 
the dust mass in the ejecta itself should
drop with time as grains are destroyed.
In the simplest picture, we thus expect that the abundances of refractory abundances
in the {\em gas-phase} supernova material itself should increase with time,
and hence be higher in older remnants. 
On the other hand,
the abundances of volatile elements should not increase as 
dramatically, so the refractory-to-volatile element
ratios observed in the gas phase in the supernova 
ejecta should be relatively low at early times, rising at
late times.  Thus we expect that such an effect
should be visible in the Fe/O ratio.

This simple picture is complicated by the sweeping up of 
interstellar dust that survived the supernova radiation.
Also, it is possible that, in some supernova remnants,
dust can form at later times, e.g., 
between the forward and reverse shocks as suggested
recently by models of \citet{sarangi2022}, and by observations of SN 1987A
that may suggest the dust mass has {\em increased} after the collision
with the circumstellar ring \citep{matsuura2019}.
Observation of supernova remnant gas-phase composition versus age would shed light on these processes,
and could give insight into dust processing by supernovae.

Spinning dust emits polarized radiation.
A nonzero net polarization in a supernova remnant would reflect an underlying ordered component to the magnetic field in the regions where the dust resides.
This is likely difficult to observe but potentially offers a probe of supernova remnant magnetism. 

\textit{Deposits in crusts and sediments.}
If and when deposits of other supernova-generated radioisotopes are found, the timescales of their pulses should also be determined by dust 
sputtering, but should in general be longer than the short timescale
expected for entrained dust. However, the timescales may vary between
different radioisotopes and may not match that of the
\fe60 deposits, due to variety in dust properties (e.g., the grain size, density, composition, initial velocity, and sputtering probability) as well as the unique geophysical cycles for different elements. The comparisons of these timescales with that for the \fe60 pulse would probe these variations. For this reason, more
data on the \pu244 observed~\citep{wallner2021} in time ranges spanning the \fe60 pulses $\sim 2.5$ and 7~Mya with better time resolution would be
particularly interesting. 
If the \pu244 signal shows two pulses each largely overlapping with the \fe60, that would suggest a common origin, while substantial \pu244 flux outside of the \fe60 pulses would indicate the the \pu244 is from another source.

\textit{Lunar \fe60.}
If the magnetic field in the ISM were negligible, the dust 
propagation would be ballistic and essentially radial. Thus the 
dust bombardment of the solar system should be as a plane wave, 
and the dust trajectories would not be deflected significantly by
either solar or terrestrial magnetic fields \citep{fry2016},
so the lunar distribution should reflect the direction of the progenitor in a clear
dependence of \fe60 with latitude \citep{fry2016}. However, the interstellar magnetic effects 
discussed above would lead to grain reflection and diffusion, leading to 
a wider distribution of solar system arrival directions. In this case,
the lunar distribution should be more homogeneous.
Clearly, further theoretical study is warranted.
A new generation of lunar sample return missions has been inaugurated by Chang'e 5,
and we strongly urge that measurements of \fe60 and other radioactive
isotopes be prioritized in this and future lunar return missions such as Artemis.
The pioneering Apollo samples were from sites relatively close to the lunar equator,
so the measurements at different lunar latitudes will be particularly interesting. We note in
this connection that the Chang'e 5 site was at a latitude $> 40^\circ$~N, and that the
planned Artemis site is close to the lunar south pole.
We also urge additional measurements of \fe60 and other radioisotopes in Apollo lunar 
samples, so as to provide a standard for comparison with the new measurements.

\textit{Presolar grains.}
Meteorites contain $\sim$ micron-sized inclusions called presolar grains, 
which manifest very different isotopic ratios from the rest of the object \citep{zinner2006, hynes2009}.  
They are understood to be interstellar dust grains that were incorporated intact in the protosolar nebula,  
and their diverse elemental and isotopic compositions point to a range of stellar sources.  
The so-called X grains are predominantly silicon carbide, but also contain
iron-peak elements and have isotopic ratios consistent with core-collapse supernovae~\citep{amari1992,kodolanyi2018}.
Some X grains indicate that live \iso{Ti}{44}
($t_{1/2} = 60 \ \rm yr$) was present at their formation \citep{nittler1996},
confirming that dust forms rapidly in supernova ejecta, and 
demonstrating that at least {\em some} supernova dust survives the remnant.  
This supports our expectation that some \fe60-bearing grains would survive 
and be successfully delivered to Earth.

\textit{Cosmic rays.}
\citet{ellison1997} 
proposed that the origin  of cosmic rays is by diffusive shock acceleration
in supernovae, and that the cosmic-ray ``seed''  ions arise from sputtered interstellar dust grains
accelerated in the remnant.  Since \fe60 has been measured in cosmic rays
\citep[see, e.g.,][]{kachelriess2015, kachelriess2018, savchenko2015, binns2016}, a detailed examination
of acceleration and sputtering of \fe60-bearing grains within the inner supernova remnant could show
a similar mechanism.  The results in Figure 5 of \citet{fry2020} show some acceleration of dust grains.  
This cosmic-ray component of \fe60 would be complementary to the refractory component, 
as it measures the sputtered component of supernova ejecta, as opposed to the surviving dust grains.  
Additionally, we expect that cosmic rays should be 
enhanced in other radioactive supernova radioisotopes of interest, e.g., \al26. 

\textit{Implications for High-Redshift Galaxies.}
The production, destruction, and survival of supernova-produced dust in galaxies are major issues in astrophysics \citep{nozawa2007,bocchio2016,micelotta2016,slavin2020}; the terrestrial \fe60 signal from 2.5 Mya offers unique insight into these processes. The \fe60 signal is detected in deep-sea deposits after
entering Earth's atmosphere in the form of dust grains, actively sampling the dust within the remnant at a specific distance from the progenitor during multiple phases of remnant evolution.  The $\gtrsim$ 1 Myr breadth of the \fe60 signal confirms that a significant amount of supernova-produced dust survives the reverse shock and for upward of 1 Myr afterwards, in order for enough of it to penetrate the solar system and fall out on the Earth and be detected by the highly sensitive AMS measurements.

The large dust content of high-redshift galaxies, now measured out to $z > 6$ \citep{lesniewska2019},
connects our work to galaxy formation and evolution. Given the young age of these systems,
it would seem that supernovae are required to produce the dust and that it survives in large quantities \citep{Todini2001,dwek2007,nozawa2007}. Indeed, dust may play a role in the galaxy outflows that are also ubiquitous at early times. \citet{squire2021} recently argued that 
charged dust is tightly coupled to cosmic rays not collisionally, but with interactions mediated by magnetic fields. 
This argument is supported by our analysis of the time span of the \fe60 signal, which is explained by these
interactions of dust particles with magnetic fields. 
These interactions could couple dust to cosmic rays, whose pressure drives outflows, and link dust to cosmic-ray confinement and escape in starburst galaxies.

\section{Conclusions} \label{sec:conc}

The deposition history of live radioisotopes on Earth has provided an opportunity to study the
propagation through the ISM of dust particles from astrophysical explosions and contribute
to resolving the issues discussed in \S\ref{sec:tests}. Specifically, the increasing maturity of
measurements of deep-sea deposits of \fe60 already enables some conclusions to be drawn, while leaving
some questions open for future studies.

\begin{itemize}
    \item To date, six papers have reported detections of \fe60 signals in deep-sea deposits that are dated to $2-4$ Mya \citep{knie2004,fitoussi2008,fimiani2016,ludwig2016,wallner2016,wallner2021}.  These papers document 13 independent samples, demonstrating that the \fe60 signal is well established and global, having been observed in all the major
    oceanic basins. 
    \item The relatively rapid growth of sediment columns, compared to FeMn crusts, permits finer sampling and better
    time resolution. In this paper, we have focused on time series analyses of the sediment data from~\cite{ludwig2016} and \cite{wallner2016}.
    For all cores, we found that the \fe60 deposition timescale is long:
    combining all of the data, our fits give 95\% confidence lower limit
    to the FW at 0.1 maximum of $\Delta t > 1.6 \ \rm Myr$.
    This is significantly longer than the timescale for the passage of a Sedov-like
    supernova blast ($\sim 0.1 \ \rm Myr$ or less).  

    \item Detailed fitting of the shape of the \fe60 signal in the sediment cores is inconclusive, as the current data do not have a well-defined onset or falloff pattern.  

    \item The current evidence does not permit any conclusion whether the observed \fe60 signal was produced by one or more supernovae. However, 
    if more than two
    supernovae could have produced the observed signals, 
    one must account for the 
    fact that the data of \citet{wallner2021} show clearly a second \fe60 signal of similar width from $\sim 7$~ Mya, with no \fe60 signals at intermediate times.
    
    \item One model that could have extended the dust raindown time is the ``pinball" model first described in \citet{fry2020}, in which ISM magnetic fields at the supernova ejecta-ISM interface confine the dust within the remnant.  In combination with the observer motion effect from the solar system's motion relative to the supernova \citep[see, e.g.,][]{chaikin2022}, these processes can account for extended delivery time.
\end{itemize}

The issues raised by our results suggest several directions for future work: 

\begin{itemize}
    \item Analyses of data from sediment cores in samples with finer time and depth resolution will help to pin down the indeterminate shape of the \fe60 signal, with data from the probable regions of onset and falloff at $\sim 3 - 4$~Mya and $\sim 1-2$~Mya, respectively, being particularly useful in this respect.     
    \item Sampling and analysis that connects the \fe60 signal observed today with the \fe60 signal from the 3 Mya pulse will contribute to our understanding on the underlying supernova dust transport mechanisms.  Specifically, the nature of the \fe60 signal from 10 kya to 1 Mya will allow us to track the solar system penetration of the dust as it is slowed and sputtered over time.  The FeMn crust measured by \citet{wallner2021} hints at this intricacy.
    \item If a distinctive shape can be found in the signal, it could provide valuable information on the astrophysics behind the dust transport mechanisms in the supernova remnant, and/or resolve the controversy whether the \fe60 signal from $\sim 3$~Mya is due to one or more supernovae.
    \item If similarly finely sliced data on other live isotopes become available, it will be interesting to compare their time structures with that of \fe60: large differences could indicate contributions from more than one supernova.
    \item In this connection, we note that the deposition of live \pu244 over a period $\lesssim 4.5$~Mya has been detected, but fine sampling of its time structure is not available. A comparison between the time structures of the \fe60 and \pu244 signals would be particularly interesting, as \pu244 is produced by the $r$-process, which is not thought to be important in most supernovae.
    \item We note also that a second \fe60 signal has been observed by \citet{wallner2021} and dated to
    $\sim 7$~Mya. When more data are available, it would be very interesting to repeat our analysis on this second signal. However, it seems prima facie also to have a width $\gtrsim 1$~Myr. A \pu244 signal from $\sim 4.5 - 9$~Mya has also been reported, but without sufficient statistics for fine time sampling.
\end{itemize}

Studies of live radioisotopes from astrophysical sources are of growing
importance in several other scientific areas beyond probing the possible
impacts on Earth of nearby supernova explosions. As discussed at length in this 
paper, the evidence of a relatively long ($\gtrsim 1$~Myr) timescale for the
deposition of \fe60 $\sim 3$~Mya is relevant for models of dust propagation,
magnetic fields in the ISM, and cosmic-ray physics. Measurements of this pulse,
together with those of the recently discovered earlier \fe60 pulse from
$\sim 7$~Mya, can cast light on the formation and evolution of the Local Bubble.
The discovery of \pu244 deposited during time periods including these \fe60
pulses may be evidence for occurrences of the $r$-process for astrophysical
nucleosynthesis in unusual supernovae and/or 
an earlier kilonova~\citep{wang2021a,wang2021b}. Measurements
of \pu244 deposition with finer time resolution will help distinguish between
the single- or multiple-supernova interpretations of the \fe60 pulse from
$\sim 3$~Mya. Finally, if a combination of a recent nearby supernova with a 
not-so-nearby, not-so-recent kilonova is implicated in the interpretation of
this pulse, it will be interesting
to continue explorations of the terrestrial impacts of earlier, closer
kilonova explosions, including their possible roles in 
mass extinctions~\citep{fields2020}.

\appendix 

\section{Fitting Functions}

\label{app:fitfuncs}

We adopt simple fitting functions for the \fe60 flux over time. 
For each function, we infer a measure of the total signal time span, which
we define as the FW at 0.1 of the maximum, which we denote by FW0.1M.
All fitting functions are for a single pulse with a unique peak.

\subsection{Three-parameter Fits}

The three parameters for all fits are $\vec{\theta} = (t_{\rm peak},\dtpar,\Phi_{\rm peak})$:
these are the time of the peak, a measure of the width, and peak flux, respectively:
\begin{eqnarray}
\mbox{Sawtooth:} \ 
\Phi_{\rm saw}(t) & = & \Phi_{\rm peak} \ \times \ 
\left\{ 
\begin{array}{cl} 
\displaystyle \frac{t-t_{\rm peak}}{\dtpar} + 1 \ &, t_{\rm peak} - \dtpar \le t < t_{\rm peak} \\[12pt]
0 \ &, \mbox{otherwise}
\end{array}
\right. \\
\mbox{Gaussian:} \
\Phi_{\rm Gau}(t) & = & \Phi_{\rm peak} \ \times \ \exp{\left[- \frac{(t-t_{\rm peak})^2}{2\dtpar^2} \right]} \\
\mbox{Triangle:} \ 
\Phi_{\rm tri}(t) & = & \Phi_{\rm peak} \ \times \ 
\left\{ 
\begin{array}{cl} 
\displaystyle \frac{t-t_{\rm peak}}{\dtpar} + 1\ &,  t_{\rm peak} - \dtpar \le t < t_{\rm peak} \\[12pt]
\displaystyle 1-\frac{t-t_{\rm peak}}{\dtpar} \ &,  t_{\rm peak} \le t < t_{\rm peak} + \dtpar \\[12pt]
0 \ &, \mbox{otherwise}
\end{array}
\right. 
\end{eqnarray}
The FW0.1M measures, $\Delta t$, of the FW are given by
\begin{eqnarray}
\mbox{Sawtooth:} \ \twidth & = &  (1-\varepsilon)\dtpar  = 0.9 \ \dtpar \\
\mbox{Gaussian:} \ \twidth & = &  
2 \sqrt{2\ln{\bfrac{1}{\varepsilon}}} \ \  \dtpar = 4.29 \ \dtpar \\
\mbox{Triangle:} \ \twidth & = &  2(1-\varepsilon)\dtpar = 1.8 \ \dtpar
\end{eqnarray}
where $\varepsilon = 0.1$ is a somewhat arbitrary choice but corresponds 
roughly to the level of the background to the measurements.

Finally, for each functional form, we calculate the corresponding fluence,
${\cal F} = \int \Phi(t) \ dt$:
\begin{eqnarray}
\mbox{Sawtooth:} \ {\cal F}_{\rm saw} & = &  \frac{1}{2} \Phi_{\rm peak} \, \dtpar \\
\mbox{Gaussian:} \ {\cal F}_{\rm Gau} & = &  \sqrt{2\pi} \ \Phi_{\rm peak} \,  \dtpar \\
\mbox{Triangle:}\  {\cal F}_{\rm tri} & = &  \Phi_{\rm peak} \, \dtpar
\end{eqnarray}

\subsection{Four-parameter Fit}

We also use an asymmetric triangle or sharktooth shape, with separate width parameters
$\sigma_{t_1}$ and $\sigma_{t_2}$ before and after the peak, respectively.  
This function allows us to probe the degree of asymmetry  or ``lean'' in the signal:
\begin{equation}
\Phi_{\rm shark}(t) = \Phi_{\rm max} \ \times \ 
\left\{ 
\begin{array}{cl} 
\displaystyle \frac{t-t_{\rm peak}}{\sigma_{t_1}} + 1\ &, t_{\rm peak} - \sigma_{t_1} \le t < t_{\rm peak} \\[12pt]
\displaystyle 1-\frac{t-t_{\rm peak}}{\sigma_{t_2}} \ &, t_{\rm peak} \le t < t_{\rm peak} + \sigma_{t_2} \\[12pt]
0 \ &, \mbox{otherwise}
\end{array}
\right. 
\end{equation}
This encompasses both our sawtooth fit ($\sigma_{t_2} = 0$)
as well as the triangle fit $(\sigma_{t_1} = \sigma_{t_2})$, 
as well as a reverse sawtooth shape.
The FW0.1M, $\Delta t$, and the fluence, $\cal F$, are defined by analogy with those for the 3-parameter triangle.

\section{Gaussian Global Fit}

\begin{figure}
    \centering
    \includegraphics[width=0.8\textwidth]{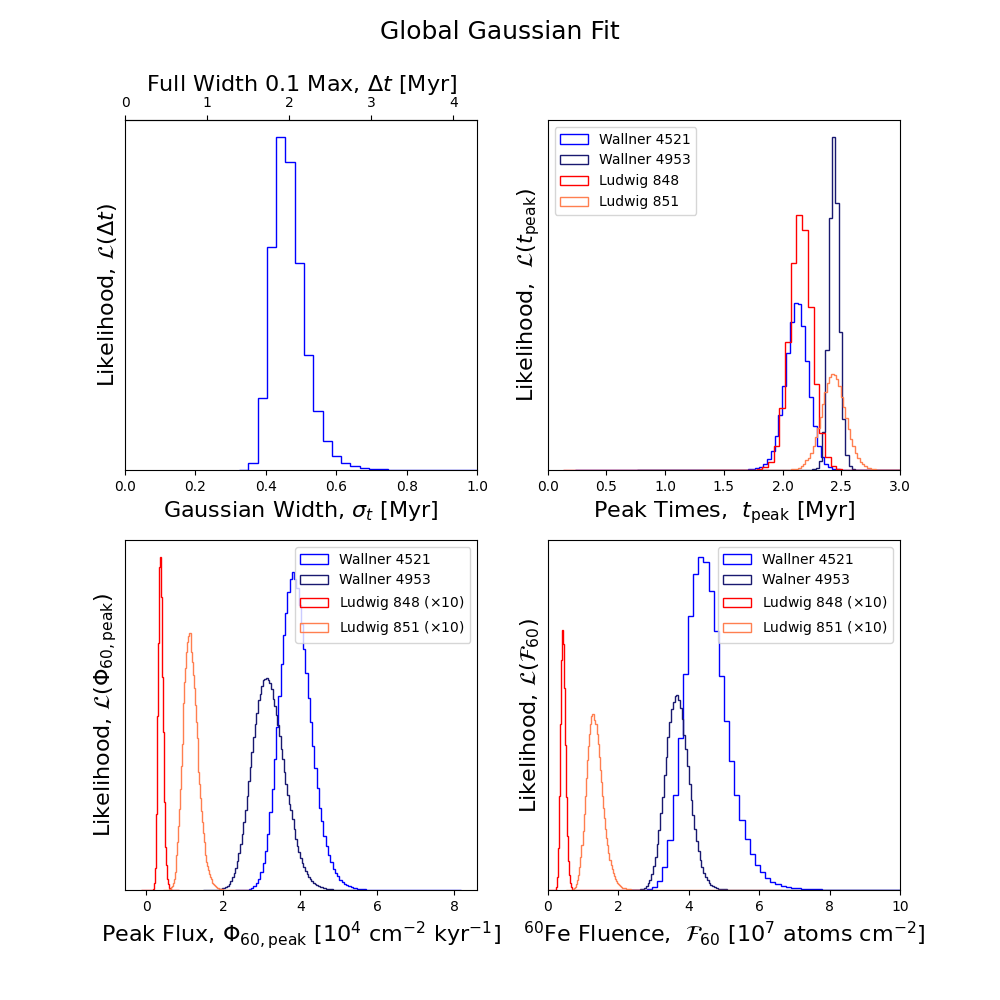}
    \caption{ {\it Global fit to all well-sampled sediments, using
    a Gaussian profile for the flux.} The width is fixed, but the peak flux and peak time are both allowed to vary, allowing for nonuniform global dust fallout and systematic errors in absolute timing, respectively. We denote the width parameter as $\dtpar$, and the full width at 0.1 maximum as  $\Delta t$.}
    \label{fig:FixW_VarPH_gauss}
\end{figure}

Figure \ref{fig:FixW_VarPH_gauss} shows the results for a global fit as in Fig.~\ref{fig:FixW_VarPH_sawtooth}, but for a Gaussian pulse shape.  Results are discussed in \S\ref{sect:global}.

\newpage

\acknowledgments
We remember Shawn Bishop as a friend
and colleague who made pioneering contributions
to this field.
We are particularly indebted to
J. Ertel and M. Goni
for illuminating discussions of geochemical, geophysical, and biological effects.
We are grateful to Shawn Bishop, Thomas Faestermann, Caroline Fitoussi, Gunther Korschinek, Peter Ludwig, and Toni Wallner
for answering our questions about their data.
It is a pleasure to acknowledge many constructive comments from the anonomyous referee, and useful discussions with
Evgenii Chaikin, Bruce Draine, Charles Gammie, Dieter Hartmann, Sasha Kaurov, Ashvini Krishnan,
Xin Liu,
Danny Milisavljevic, Jesse Miller, Paul Ricker, Danylo Sovgut, Rebecca Surman, Alexandra Trauth, and Xilu Wang.
B.D.F. is grateful for fruitful discussions with
all of the participants of the ``Historical Supernovae, Novae, and Other Transients'' workshop held in 2019 October, and to the Lorentz Center at the University of Leiden for their hospitality in hosting this event.
The work of A.F.E. and B.D.F. was supported in part by the NSF under grant number AST-2108589, and benefited from grant No.~PHY-1430152 (JINA Center for the Evolution of the Elements).
The work of J.E. was supported by the United Kingdom STFC grants ST/P000258/1 and ST/T000759/1,
and by the Estonian Research Council via a Mobilitas Pluss grant.

\bibliography{References2.bib}

\end{document}